\documentclass[referee]{aa}
\usepackage[latin1]{inputenc}
\usepackage{natbib}
\usepackage{theorem,algorithm,algorithmic}
\usepackage{amssymb,amsmath}
\usepackage{placeins}

\input cyracc.def
\font\tencyr=wncyr10
\def\cyr{\tencyr\cyracc}

\usepackage{psfig,epsfig,amssymb,alltt}
\usepackage{latexsym}
\usepackage{url}
\usepackage{graphicx}
\graphicspath{{PS/}}
\usepackage{psfig}
\usepackage{epsfig}
\psfigurepath{./PS:.}
\usepackage[english]{babel}
\usepackage{indentfirst}
\psfigurepath{./PS}
\vspace{1cm}

\begin{document}
\title{Gap interpolation by inpainting methods : \\ Application to Ground and Space-based Asteroseismic data}
 \author{Sandrine Pires\inst{1}  \and Savita Mathur\inst{1, 2, 3} \and Rafael A. Garc\'ia\inst{1} \and J\'er\^ome Ballot \inst{4,5} \and Dennis Stello\inst{6} \and Kumiko Sato\inst{1}}

 \institute{Laboratoire AIM, CEA/DSM-CNRS, Universit\'e Paris 7 Diderot, IRFU/SAp-SEDI, Service d'Astrophysique, CEA Saclay, Orme des Merisiers, 91191 Gif-sur-Yvette, France\\
     \email{sandrine.pires@cea.fr}
	\and High Altitude Observatory, NCAR, P.O. Box 3000, Boulder, CO 80307, USA
	\and Space Science Institute, 4750 Walnut street Suite \#205, Boulder, CO 80301, USA
	\and CNRS Institut de Recherche en Astrophysique et Plan\'etologie, 14 avenue Edouard Belin, 31400 Toulouse, France
	\and Universit\'e de Toulouse, UPS-OMP, IRAP 31400, Toulouse, France
	\and Sydney Institute for Astronomy (SIfA), School of Physics, University of Sydney, NSW 2006, Australia
             }

\offprints{sandrine.pires@cea.fr}

\date{\today}


\abstract{\\
In asteroseismology, the observed time series often suffers from incomplete time coverage due to gaps. The presence of periodic gaps may generate spurious peaks in the power spectrum that limit the analysis of the data.
Various methods have been developed to deal with gaps in time series data. However, it is still important to improve these methods to be able to extract all the possible information contained in the data. 
 In this paper, we propose a new approach to handle the problem, the so-called inpainting method. This technique, based on a sparsity prior, 
enables to judiciously fill-in the gaps in the data, preserving the asteroseismic signal, as far as possible. The impact of the observational window function is reduced and the interpretation of the power spectrum is simplified.
This method is applied both on ground and space-based data. It appears that the inpainting technique improves the oscillation modes detection and estimation. Additionally, it can be used to study very long time series of many stars because its computation is very fast. For a time series of 50 days of CoRoT-like data, it allows a speed-up factor of 1000, if compared to methods of the same accuracy.}
\maketitle 
\markboth{Inpainting for Asteroseismic data}{}


\section{Introduction}

Asteroseismology, the study of stellar oscillations, has proved to be a very powerful tool to probe the internal structure of stars.
Oscillation modes can be used to obtain reliable information about stellar interiors and this has motivated a huge observational effort with the aim of doing asteroseismology.

Until recently, ground-based observations have been the primary source of our knowledge about oscillation modes in different types of pulsating stars.
In the last decade, missions like WIRE \citep[Wide-field InfraRed Explorer,][]{wire:buzasi04}, MOST \citep[Microvariability and Oscillations of STars,][]{most:walker03}, CoRoT \citep[Convection, Rotation and Planetary Transits, ][]{corot:baglin06} or {\it Kepler} \citep{kepler:borucki10} have made it possible to simultaneously record long time series on numerous targets. 
Nevertheless, ground-based networks such as SONG \citep[Stellar Oscillations Network Group, ][]{song:grundahl06} can still be used in a complementary way to observe stars much brighter than the ones observed from space.
All these missions have allowed us to study the internal structure of a lot of stars \citep[e.g.][]{JCDDap1996,kepler:mathur12,corot2:mathur13} their dynamics, including internal rotation \citep[e.g.][]{sun:fletcher06,2012Natur.481...55B,kepler:deheuvels12,kepler:mosser12,kepler:gizon13}, their magnetism \citep[e.g.][]{corot:garcia10,kepler:karoff13} and their fundamental parameters such as their masses, radii, and ages \citep[e.g.][]{kepler:mathur12,kepler:metcalfe12,kepler:dogan13,corot:mathur13,astero:chaplin13}

However, to be able to extract all the possible information on the stochastically excited oscillations of  solar-like stars from the light curves, it is important to have continuous data without regular gaps. 
One important criterion for observational asteroseismology is the duty cycle. This is the measure of the fraction of time that is spent successively observing the variability of a given star. From the ground, for extended periods, even when instruments (or telescopes) from the same networks are deployed at different longitudes, the day/night cycle, the weather and other factors make it impossible to obtain continuous data sets. Normal duty cycles of ground based networks of several continuous months are typically below 90\%.
From space, most of these problems are overcome and duty cycles are commonly above 90\%. However, even small time gaps can cause significant confusion in the power spectrum. For example, the presence of repetitive gaps induced when a spacecraft, in a low Earth orbit, crosses the so-called South Atlantic Anomaly, may generate spurious peaks in the power spectrum, as it is the case for the CoRoT satellite. 

In helioseismology of Sun-as-a-star instruments  GOLF \citep[Global Oscillations at Low Frequency, ][]{golf:gabriel95} and VIRGO \citep[Variability of Solar Irradiance and Gravity Oscillations, ][]{virgo:frohlich95} no interpolation is used. Missing points are filled with zeros assuming that the time series have been properly detrended and thus have a zero mean \citep{golf:garcia05}. In particular, for the SoHO instruments, the average duty cycle is around $95\%$ and the gaps are concentrated during the SoHO vacation periods \citep[see more details about the GOLF data in][]{golf:garcia05}. In the case of BiSON \citep[Birmingham Solar-Oscillations Network, ][]{bison:chaplin96}, only 1-point gaps are filled using a cubic spline. For imaged instruments --such as GONG \citep[Global Oscillation Network Group, ][]{gong:harvey96} and MDI \citep[Michelson Doppler Imager, ][]{mdi:scherrer95} -- only gaps of one or two images are filled using autoregressive algorithms  \citep[e.g.][]{AR:fahlman82}. Recently, in the case of HMI \citep[Helioseismic and Magnetic Imager, ][]{hmi:scherrer12}, gaps up to a certain size are corrected \citep{hmi:larson08}. Autoregressive algorithm could be a possible choice in asteroseismology but they need to be further studied and tested for other stars and data sets.

In asteroseismology, it is common to directly compute the periodogram using the light curves with gaps using algorithms such as the Lomb-Scargle periodogram \citep{kepler:lomb76,kepler:ferraz81,kepler:scargle82,kepler:frandsen95,kepler:zechmeister09} or CLEAN  \citep{kepler:roberts87,kepler:foster95}. It is also commonly used to interpolate the missing data to estimate the power spectrum with a Fast Fourier Transform.
In particular, this is the case for the seismic light curves provided by the CoRoT mission in which a linear interpolation is performed in the missing (or bad) points \citep[e.g.][]{2008A&A...488..705A,2009A&A...506...41G,corot:auvergne09}. However, in some cases a better interpolation algorithm has been used in the analysis of the CoRoT data \citep[e.g.][]{corot:mosser09,fit:ballot11}. It is important to note that the light curves have been properly detrended and have a zero mean \citep[e.g.][]{kepler:garcia11}.

The aim of this paper is to present a new interpolation method to reduce the undesirable effects  on the power spectrum coming from incomplete time coverage. We want to reduce, as much as possible, the presence in the power spectrum of non desirable peaks due to the absence of data, without modifying the seismic signals from the stars. For this purpose, we propose a new approach, the so-called inpainting technique, based on a sparsity prior introduced by \cite{kepler:elad05}. The method we introduce in this paper, consists of filling the gaps prior to any power spectrum estimation. 

In section 2, we first recall the basis of the classic methods currently used in asteroseismology. Then, the inpainting technique is presented. 
In section 3, we study the impact of inpainting techniques in the power spectrum using solar-like data on which we apply standard window functions from ground-based observations, as well as other artificially-modeled masks. Then, in section 4, a similar study is performed but this time using standard windows from space-based observations.
 Finally, our conclusions are summarized in section 5.

\section{Data Analysis}

To measure with a good precision the characteristics of the acoustic modes (frequencies, amplitudes, lifetimes...) in solar-like stars, long and uninterrupted light curves are desired. Both ground and space observations have gaps that can be of different lengths. They can introduce spurious peaks in the power spectrum, especially if they are regularly distributed.

\subsection{Power spectrum estimation}

Various methods have been developed to deal with incomplete time series. 

\subsubsection{Lomb-Scargle periodogram}
One of the most common methods used is the Lomb-Scargle periodogram introduced by \cite{kepler:lomb76,kepler:scargle82} which is based on a least-squares fitting of sine waves of the form  $y = a\cos wt + b \sin wt$. 
Another variant, described in \cite{kepler:frandsen95}, consists of using statistical weights to reduce the noise in the power spectrum at low frequencies.
However, these strategies are subject to false detections due to the observational window function. 

\subsubsection{CLEAN algorithm}
Another widely used method is the CLEAN algorithm \citep{kepler:roberts87,kepler:foster95} that is based on an iterative procedure that searches for maxima in the power spectrum. 
This procedure consists of finding the highest peak in the periodogram, then to remove it in the time domain, recompute the power spectrum and iterate for the following highest peak. This procedure is iterated until the residual power spectrum is a pure noise spectrum. 
At each iteration the highest peak is removed but also all the spurious frequencies that arises from spectral leakage.
By this method, false peaks can be removed.
However, any error on the properties of the peaks to be removed (amplitude, frequency, and phase) will introduce significant errors on the resulting "cleaned" periodogram.
This is especially true when the duty cycle is low and the spectral leakage of the main lobe into the side lobes is very large.

\subsection{Gap filling and regular sampling}
The detection and estimation of oscillation modes in the power spectrum of an incomplete time series sometimes suffer false detection due to large observational gaps. 
In order to lower the impact of the gaps on the estimation of the power spectrum, specific methods have already been developed to interpolate the missing data in order to estimate the power spectrum with a Fast Fourier Transform. 

In some cases, a linear interpolation is sufficient to do so \citep[e.g.][]{2008A&A...488..705A,2009A&A...507L..13B,2009A&A...506...41G,kepler:deheuvels10}, but in other cases a more sophisticated algorithm is necessary \citep[e.g.][]{corot:mosser09}.
In this paper, we present a new approach that consists of filling in the gaps using a sparsity prior. We note that there are other interpolation algorithms, such as autoregressive models, that could also be applied to asteroseismic data (as it is for example the case in helioseismology for the HMI pipeline, Larson et al. (in preparation)). Because these algorithms are not commonly used in asteroseismology, we have not shown any comparison with them in the present paper.

\subsubsection{Inpainting introduction}

A solution that has been proposed to deal with missing data consists of filling-in judiciously the gaps by performing an ``inpainting" method. The inpainting technique is an extrapolation of the missing information using some priors on the solution. This technique has already been used to deal with missing data for several applications in astrophysics \citep{kepler:abrial08,kepler:pires09} including asteroseismology \citep{inpainting:sato10}. 
In these applications the authors use inpainting introduced by \cite{kepler:elad05}. This inpainting rely on a prior of sparsity which can be easily applied to asteroseismic data.
Some methods in asteroseismology have already used the fact that the spectra are sparse. For example, in the CLEAN algorithm, a sparsity prior is introduced implicitly. In autoregressive methods such as in \cite{AR:fahlman82}, a sparsity prior is also used but the degree of sparsity is imposed in advance; it requires to have some priors on the number and the frequency range of the oscillations modes. 

The method presented here just uses the prior that there exists a representation $\Phi^T$ of the time series $X(t)$ where most coefficients $\alpha = \Phi^TX$  are close to zero. 
For example, if the time series $X(t)$ was a single sine wave, the representation $\Phi^T$ would be the Fourier transform because most of the Fourier coefficients $\alpha$ are equal to zero except one coefficient that is sufficient to represent the sine wave in the Fourier space.

Let $X(t)$ be the ideal complete time series, $Y(t)$ the observed time series (with gaps) and $M(t)$ the binary mask (i.e. window function with $M(t)=1$ if we have information at data point $X(t)$, $M(t)=0$ otherwise). We have: $Y=MX$. So, when M(t) = 0, the observed value (e.g. a flux or radial velocity measurement) is set to zero, rather than removing the data point.
Inpainting consists of recovering $X$ knowing $Y$ and $M$. 
There is an infinite number of time series $X$ that can perfectly fit the observed time series $Y$. Among all the possible solutions, we search for the sparsest solution in the representation $\Phi^T$ (i.e. the time series $X$ that can be represented with the smallest number of coefficients $\alpha$, such as the sine curve in the Fourier representation) while imposing that the solution is equal to the observed data within the intrinsic noise of the data.
Thus, the solution is obtained by solving:
\begin{equation}
\min  \| \alpha \|_0    \textrm{ subject to }  \parallel Y - MX   \parallel^2 \le \sigma,
\label{functional}
\end{equation}
where the $l_0$ pseudo-norm  $\| \alpha \|_0$ is the number of non-zero coefficients $\alpha$ and $|| . ||$ is the classical $l_2$ norm (i.e. $|| z || =\sum_k (z_k)^2$), and $\sigma$ is the standard deviation of the noise in the observed time series \citep[for further details please see][]{kepler:pires09}.

It has been shown by \cite{kepler:donoho_01} that the $l_0$ pseudo-norm can be replaced by the convex $l_1$ norm (i.e. $ || z ||_1 = \sum_k | z_k |$) if the time series $X$ is sparse enough in the representation $\Phi^T$ (i.e. a few large coefficients can represent the data). This representation is described in the next section. Thus, its global minimum can be reached by decent techniques. 

\subsubsection{Description of the algorithm}
The solution of such an optimization task can be obtained through an iterative algorithm called Morphological Component Analysis (MCA) introduced by \cite{kepler:elad05}. 
 Let $X^i$ denote the reconstructed time series at iteration $i$.
 If the time series is sparse enough in the representation $\Phi^T$, the largest coefficients should originate from the signal we want to measure.
Thus, the algorithm is based on a threshold that decreases exponentially (at each iteration) from a maximum value to zero. 
By accumulating more and more high coefficients through each iteration, the gaps in $X^i$ are filling up steadily and the power of the coefficients due to the gaps is decreasing.
This algorithm needs as inputs the observed incomplete data $Y$ and the binary mask $M$.

The algorithm can be described as follows:
\begin{enumerate}
  \item Set the maximum number of iterations $I_{max}$, the solution $X^0$ is initialized to zero, 
    the maximum threshold $\lambda_{max} = \max(\mid \Phi^T Y \mid)$, and the minimum threshold $\lambda_{min} << \sigma$.
    \item Set $i = 0$, $\lambda^0 = \lambda_{max}$. Iterate:
    \item Set $U^i = X^i + M(Y-X^i)$ to enforce the time series to be equal to the observed data where the mask is equal to $1$.
    \item Compute the forward transform of $U^i$: $\alpha = \Phi^TU^i$.
    \item Compute the threshold level $\lambda^i = F(i,\lambda_{max},\lambda_{min})$, where $F$ is a function that describes the decreasing law of the threshold.
    \item Keep only the coefficients $\alpha$ above the threshold $\lambda^i$.
    \item Reconstruct $X^{i+1}$ from the remaining coefficients $\tilde \alpha$ : $X^{i+1} = \Phi \tilde\alpha$.
    \item Set $i=i+1$. If $i<I_{max}$, return to step 3.
\end{enumerate}

As mentioned previously,  in this application, $F$ decreases exponentially from $\lambda_{max}$ at the first iteration, to $\lambda_{min}$ at the last iteration. 
When $\lambda_{min}$ is much smaller than the noise level, $\sigma$, and thus very close to zero, the algorithm assigns it the value of zero to avoid numerical issues. Setting $\lambda_{min}$ to zero when the threshold is very small is not really affecting the results because we are well below the noise level $\sigma$. Note also that the threshold can be stopped at 2 or $3\sigma$ if one wants to denoise the time series.

The number of iterations $I_{max}$ is chosen to be $100$, which ensures convergence. This value is obtained experimentally, and is not specific to asteroseismic data \citep{kepler:pires09}.
The conditions under which this algorithm provides an optimal and unique sparse solution to Eq. \ref{functional} have been explored by a number of authors \citep[e.g.][]{kepler:elad02,kepler:donoho03}. They have shown that the proposed method is able to recover the sparsest solution provided this solution is indeed sparse enough in the representation $\Phi^T$ and the mask is sufficiently random in this representation.
 In asteroseismology, $\Phi^T$ can be represented by a Multiscale Discrete Cosine Transform. Like the Fourier transform, the Discrete Cosine Transform (DCT) is a decomposition into a set of oscillating functions and thus it is a good representation of the asteroseismic signal. To go one step further and treat the large variation of gap sizes present in helio- and asteroseismic data, the light curves were decomposed beforehand with an ``\`a trous" wavelet transform \citep[see][]{book:starck02} (using a $B^3$-spline scaling function). Then, each wavelet plane is decomposed using a local DCT whose block size $B_l$ depends on the scale of the wavelet plane $l$ as follows: $B_l = B_0*2^l$. This corresponds to what we call the Multiscale Discrete Cosine Transform.


\section{Application to low duty cycle time series}
\label{ground}


Ideally, we would like to have a 100\% duty cycle when analyzing time series. 
Unfortunately, from a single ground-based site, we can only reach duty cycles of about 40\% for short time series under ideal weather conditions.  For long time series, duty cycles are generally between 20 and 30\%.
Combining several sites leads to duty cycles between 60\% and 95\% depending on the number of sites and their position in terms of longitude.

The time series obtained from the ground are mainly affected by the day/night cycle and weather conditions. 

Thus we can define a toy model where the window function (i.e. mask) is modeled by a rectangular window $\Pi_{T_1}$ convolved by a Dirac comb function ${\mbox{\cyr Sh}}_{T_2}$, with $T_1$ corresponding to the daily observation window that is about 6 hours and $T_2$ corresponding to the Earth rotation period of 24 hours.
In the Fourier domain, the multiplication by this mask becomes a convolution, and then the Fourier transform of the signal is convolved by the Fourier transform of the mask $\hat M(f)$:

\begin{equation}
\hat M(f) = T_1 \frac{\text{sin}(\pi f T_1)}{\pi f T_1} \cdot \frac{1}{T_2} {\mbox{\cyr Sh}}_{\frac{1}{T_2}}(f).
\label{specwind}
\end{equation}

The product of a signal and a Dirac comb is equivalent to a sampling. 
Thus, the convolution by a regular window function causes a spectral leakage from the main lobe to the side lobes that are regularly spaced in frequency by $\frac{1}{T_2}$ and their heights is modulated by the sinc term. This window effect is depending both on $T_1$ and $T_2$. 


For this study, we use the time series obtained by the SunPhotometers (SPM) of VIRGO \citep[Variability of Solar Irradiance and Gravity Oscillations, ][]{virgo:frohlich95} instrument onboard the SoHO spacecraft (with a duty cycle close to 100\%) in order to mimic ground-based observations. We have only considered 50 days of observations in order to be close to what we expect for a typical long observation run with the SONG network. This is our ideal time series $X(t)$.
\begin{figure}[htp]
\centerline{
\includegraphics[width=9.cm]{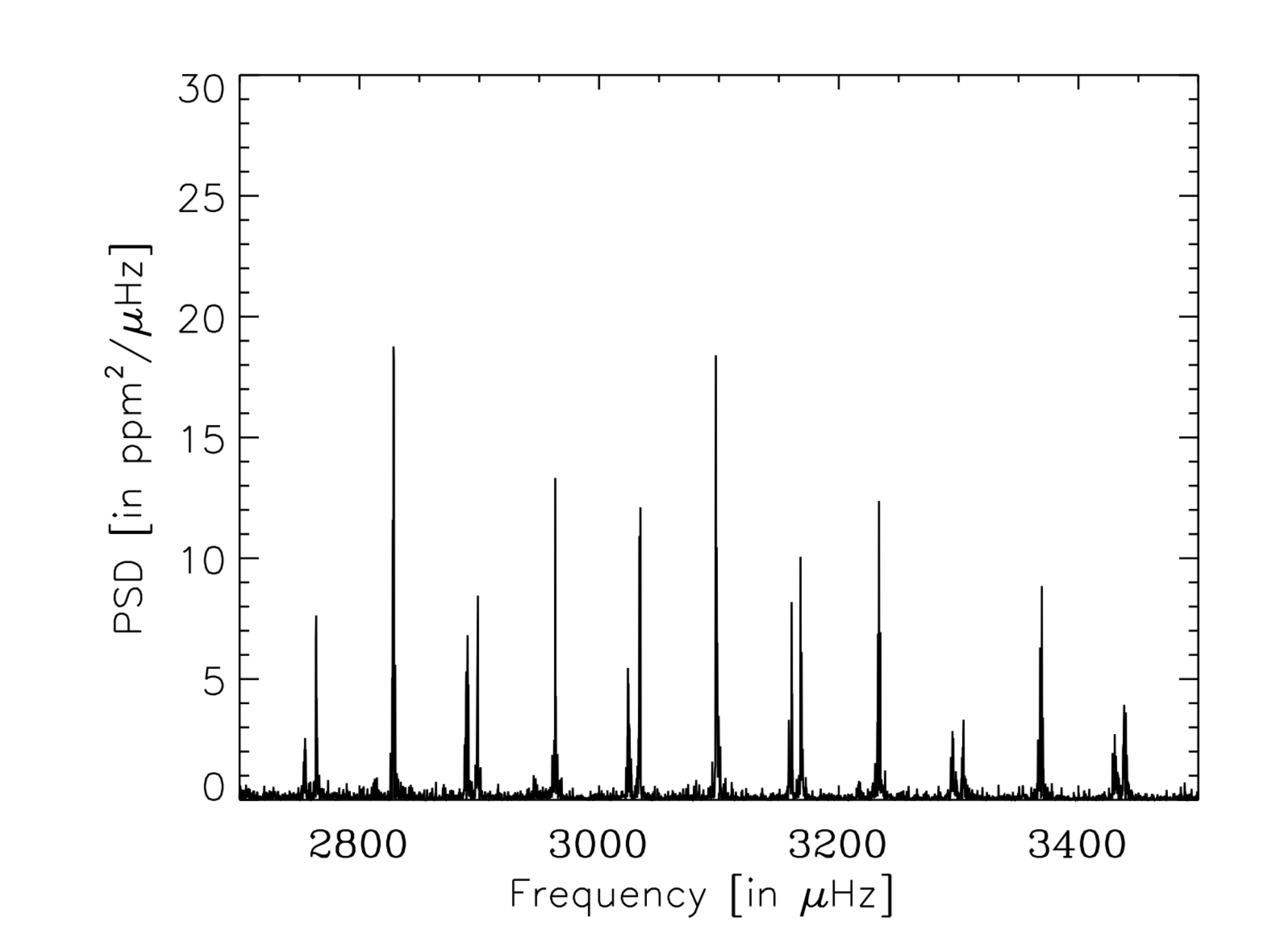}
}
\centerline{
\includegraphics[width=9.cm]{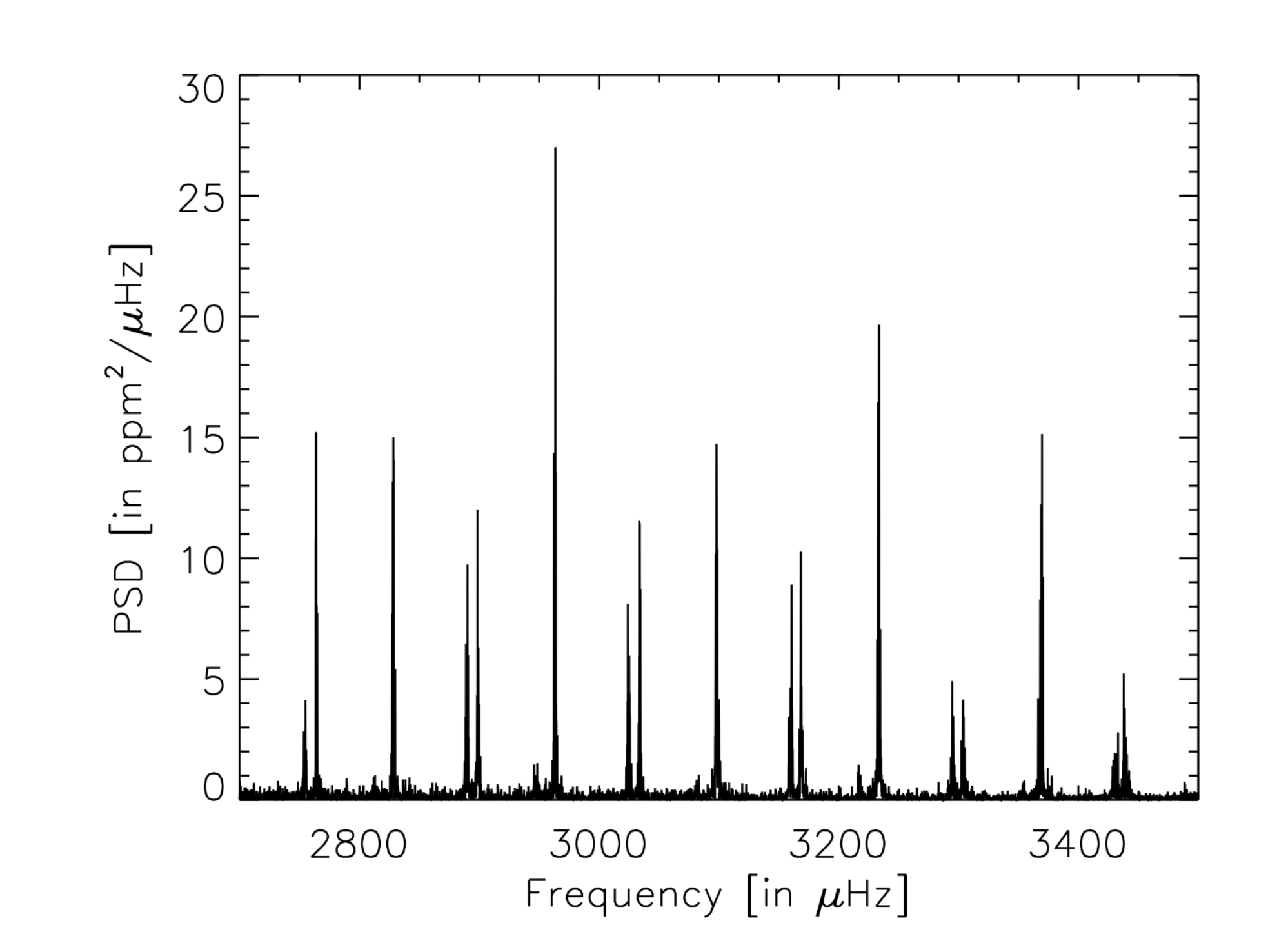}
}
\caption{Power Density Spectrum (in units of ppm$^2/\mu$Hz) of VIRGO/SPM data in the frequency range between $2700$ to $3500$ $\mu$Hz estimated with a Fast Fourier Transform from a complete time series of 50 days (top) and 91.2 days (bottom).}
\label{original_psd}
\end{figure}
The top panel of Fig. \ref{original_psd} shows the Power Density Spectrum (PDS) estimated from 50 days observations with a complete time coverage (duty cycle of 100\%).
We have only plotted the PDS in the frequency range between $2700$ to $3500$ $\mu$Hz to focus around the maximum power of the p modes.
For this simple case, with regularly sampled data, without gaps, the spectrum was computed using a Fast Fourier Transform
and we normalized it as the so-called one-sided PDS \citep{num:press92}.

\subsection{Ground-based observations of single- to multiple-site networks: worst-case scenario}

We start by studying the impact of different standard window functions from ground-based observations that have different duty cycles. To do so, we apply three different window functions to the VIRGO/SPM data. 

\begin{figure}[htp]
\centerline{
\includegraphics[width=9cm]{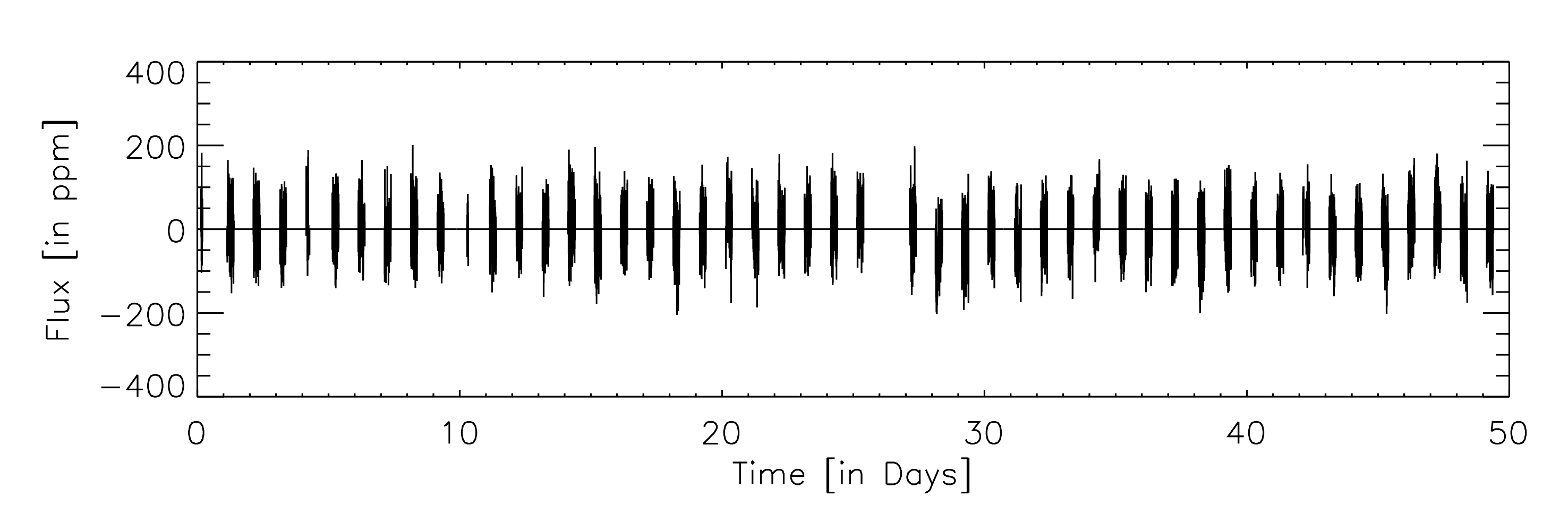}
}
\centerline{
\includegraphics[width=9cm]{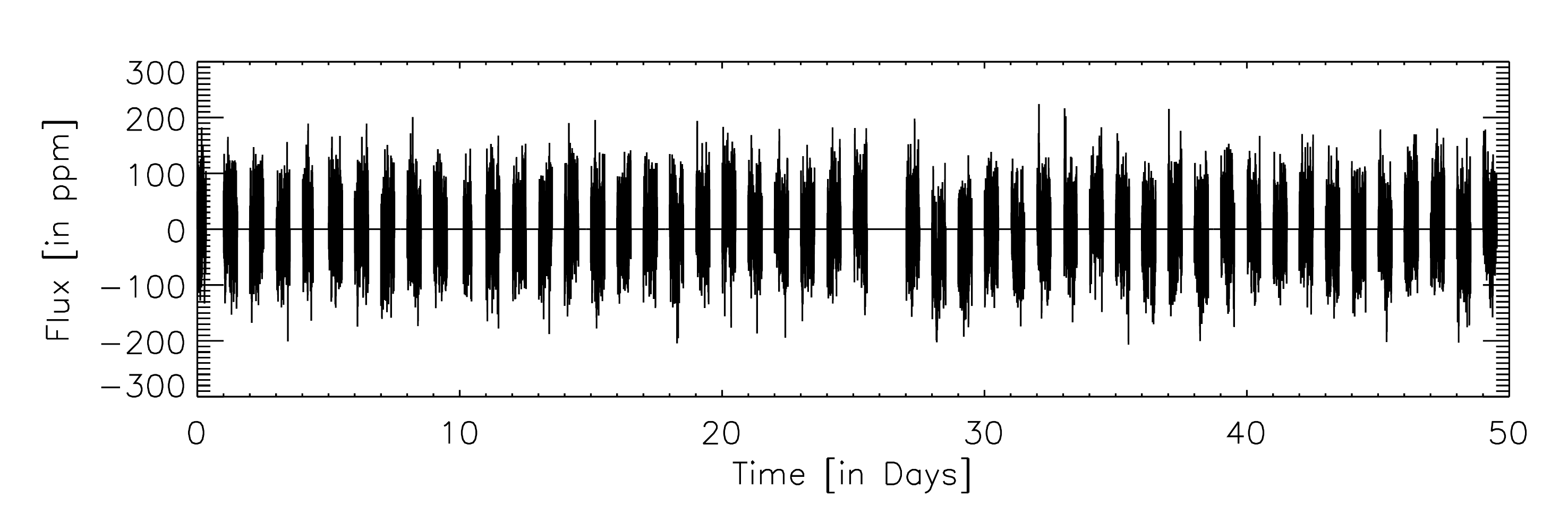}
}
\centerline{
\includegraphics[width=9cm]{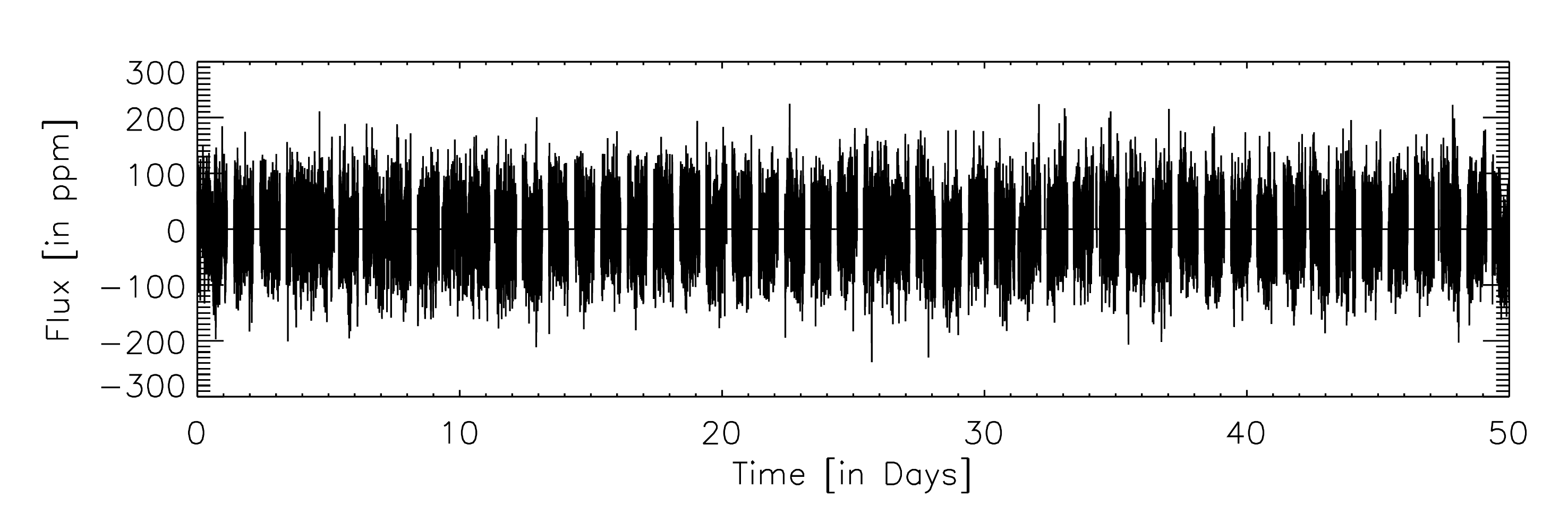}
}
\caption{Sample of 50 days of the original VIRGO/SPM data multiplied by a MARK-I like mask to simulate ground-based observations from: a single site corresponding to a duty cycle of 23\% (top panel), two different sites with a duty cycle of 50\% (middle panel), and from multiple sites corresponding to a duty cycle of 77\% (bottom panel).}
\label{dc23_original}
\end{figure}

To study the case of ground-based observations from a single site, we have expanded the MARK-I window function to simulate a duty cycle of 23\%. 
MARK-I is one of the helioseismic Doppler-velocity instruments of the BiSON network operated at the Observatorio del Teide, Tenerife \citep{mark1:chaplin98} and located close to the first SONG node. 
This window function is therefore a reasonable approximation of what the observations with one node will have during its operation.
We have applied a dilatation operator (i.e. an operator that enlarges the regions were information is provided) to the MARK-I window function to simulate the case of ground-based observations from two different sites with a duty cycle of 50\%. Although this mask is not typical for real data, it contains highly periodic gaps. This allows us to test our algorithm in the worst-case scenario and to establish a lower limit of any improvement of the inpainting algorithm compared to CLEAN.
Finally, a mask for a multiple-site network is obtained by inverting the MARK-I window function providing a duty cycle of 77\%. 
Although, day-time and night-time observations are not the same and instruments are different, these artificial regular masks are sufficiently representative for ground-based network observations for us to reasonably compare the different methods, these artificial regular masks represent the worst-case scenario for ground-based observations, because they have a high number of regular gaps.
The time series $X$ is then multiplied by these observational window functions $M$ to obtain the artificial observed data $Y$ (see Fig. \ref{dc23_original}).

These sets of data were then analyzed with the three methods described in section 2.
A comparison with linear interpolation for gap filling was shown in previous work, based on CoRoT data, which concluded that the inpainting method was superior \citep{inpainting:sato10}. Nevertheless, we have applied the linear interpolation to the sets of simulated ground-based data described above and the results are very poor compared to the results obtained with the three methods described in this paper. 

\begin{figure}[htp]
\centerline{
\includegraphics[width=9cm, height=5.9cm]{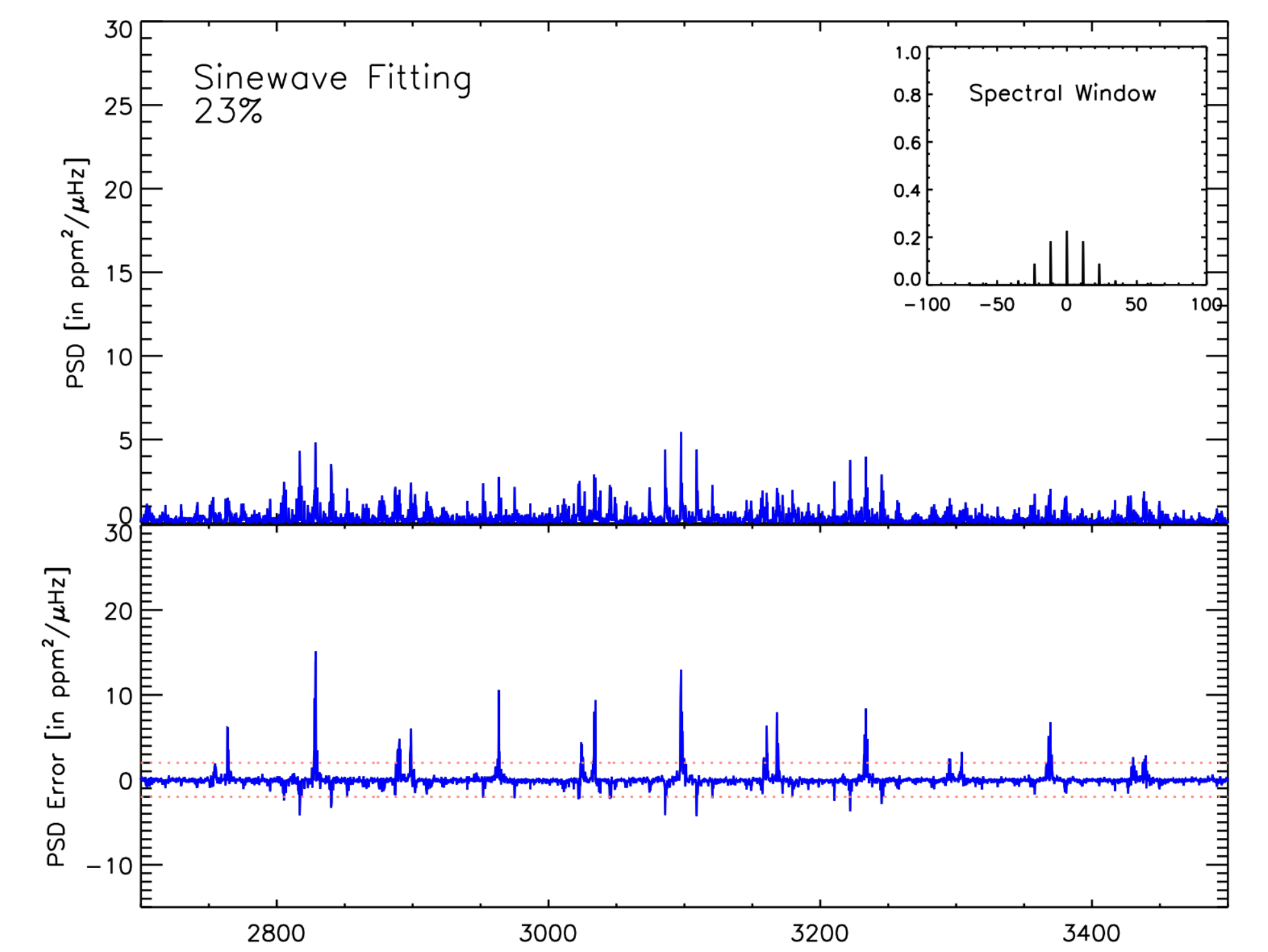}
}
\vspace{0.cm}
\centerline{
\includegraphics[width=9cm, height=5.9cm]{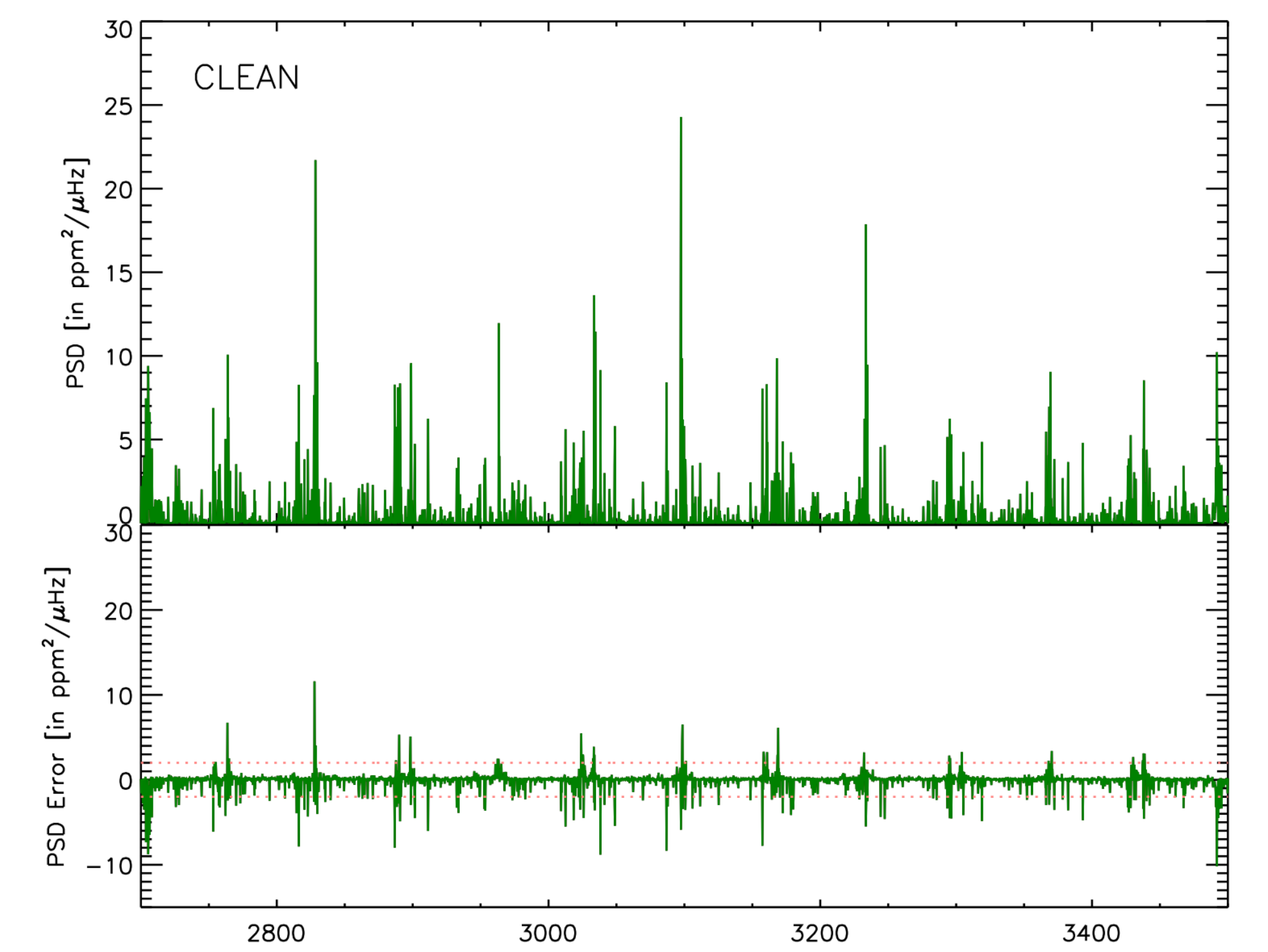}
}
\centerline{
\includegraphics[width=9cm, height=6.3cm]{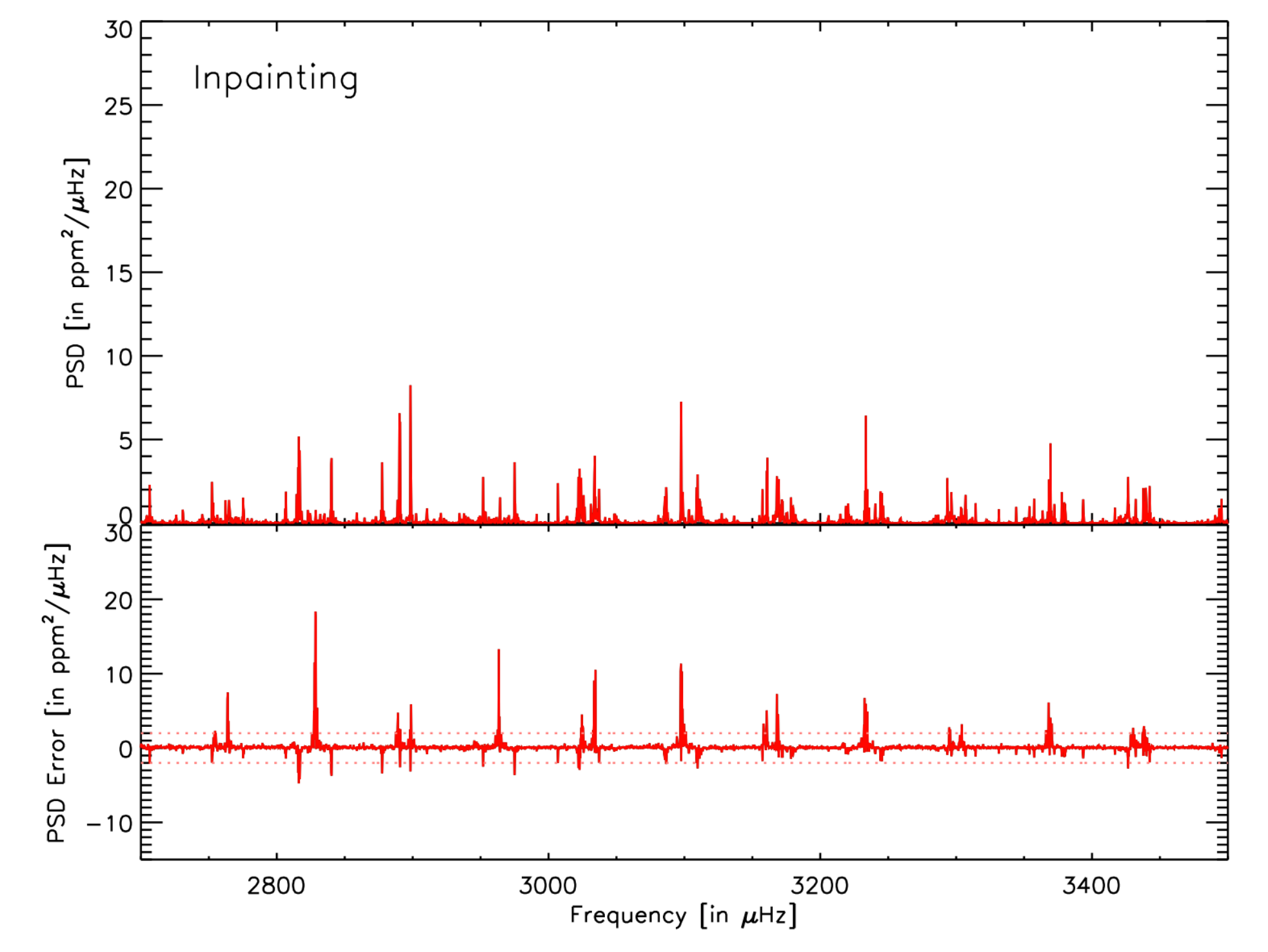}
}
\caption{Power Density Spectrum (in units of ppm$^2/\mu$Hz) for a duty cycle of 23\%. The PDS is calculated with a least-squares sinewave fit on the incomplete time series (top panel), using an FFT on the CLEANed time series (middle panel) and using an FFT on the inpainted time series (bottom panel). The errors correspond to the difference between the original PDS and the PDS estimated from the incomplete data. The inset corresponds to the spectral window of the time series.}
\label{dc23}
\end{figure}

\begin{figure}[htp]
\centerline{
\includegraphics[width=9cm, height=5.9cm]{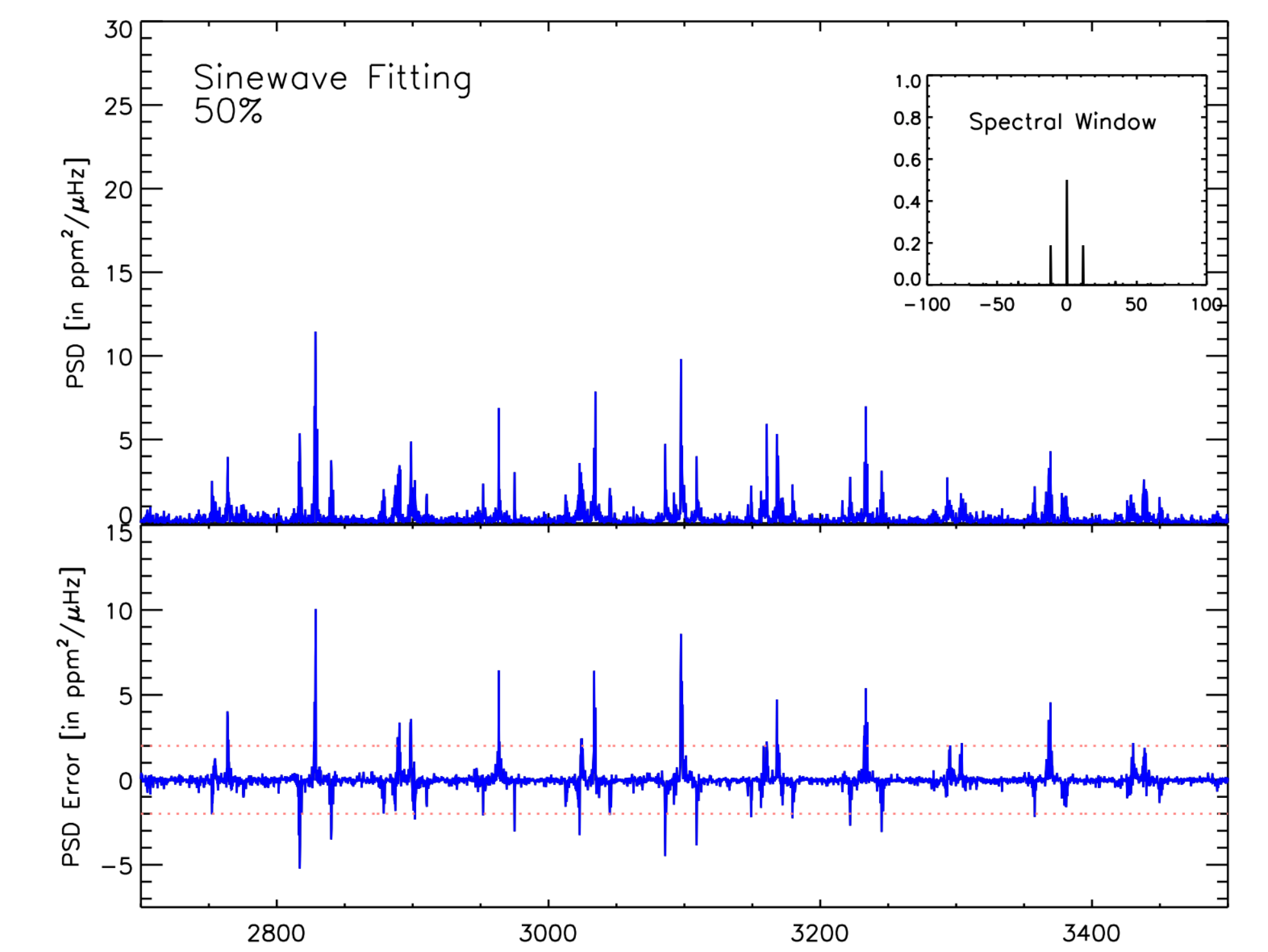}
}
\vspace{0.cm}
\centerline{
\includegraphics[width=9cm, height=5.9cm]{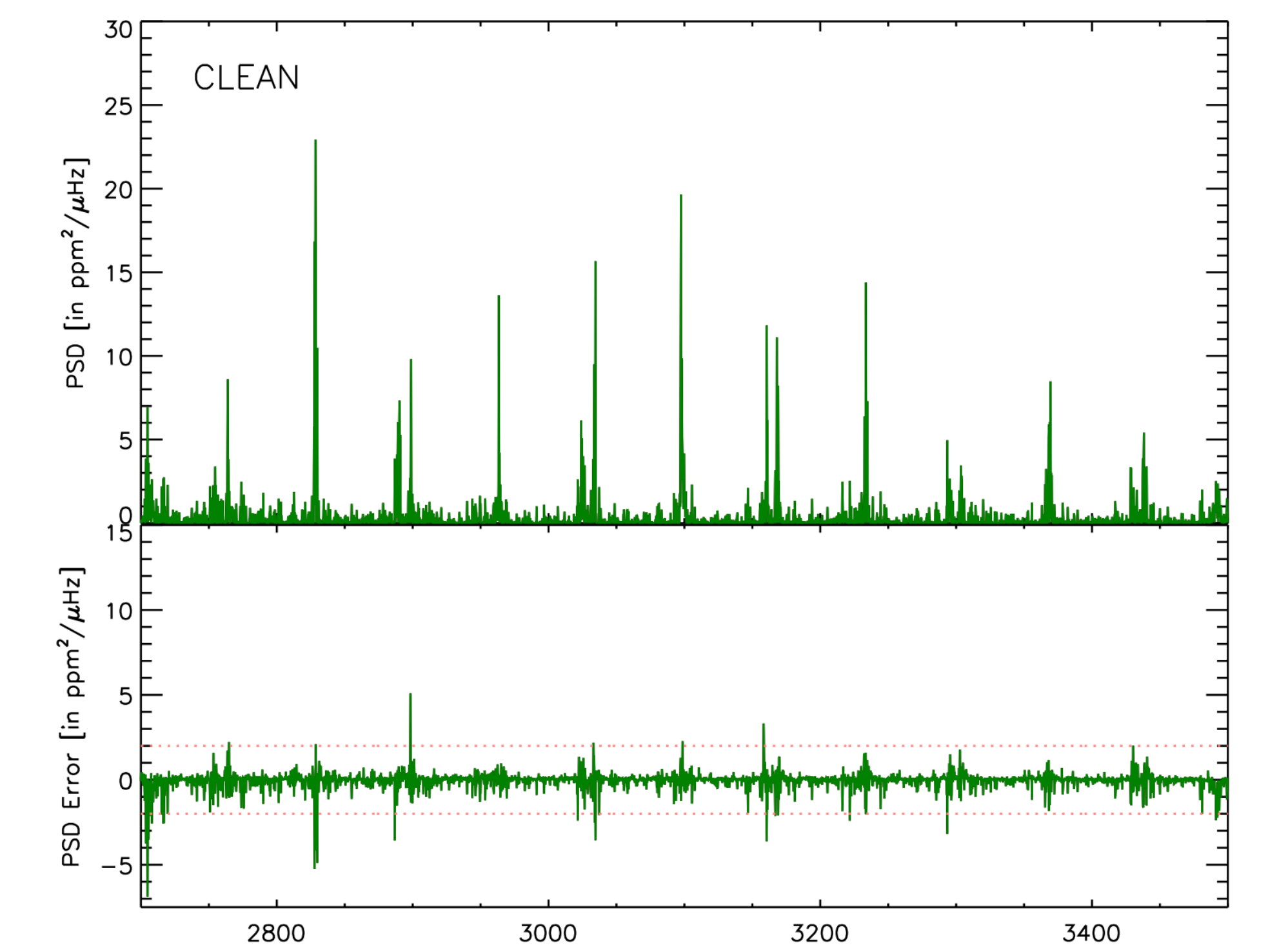}
}
\centerline{
\includegraphics[width=9cm, height=6.3cm]{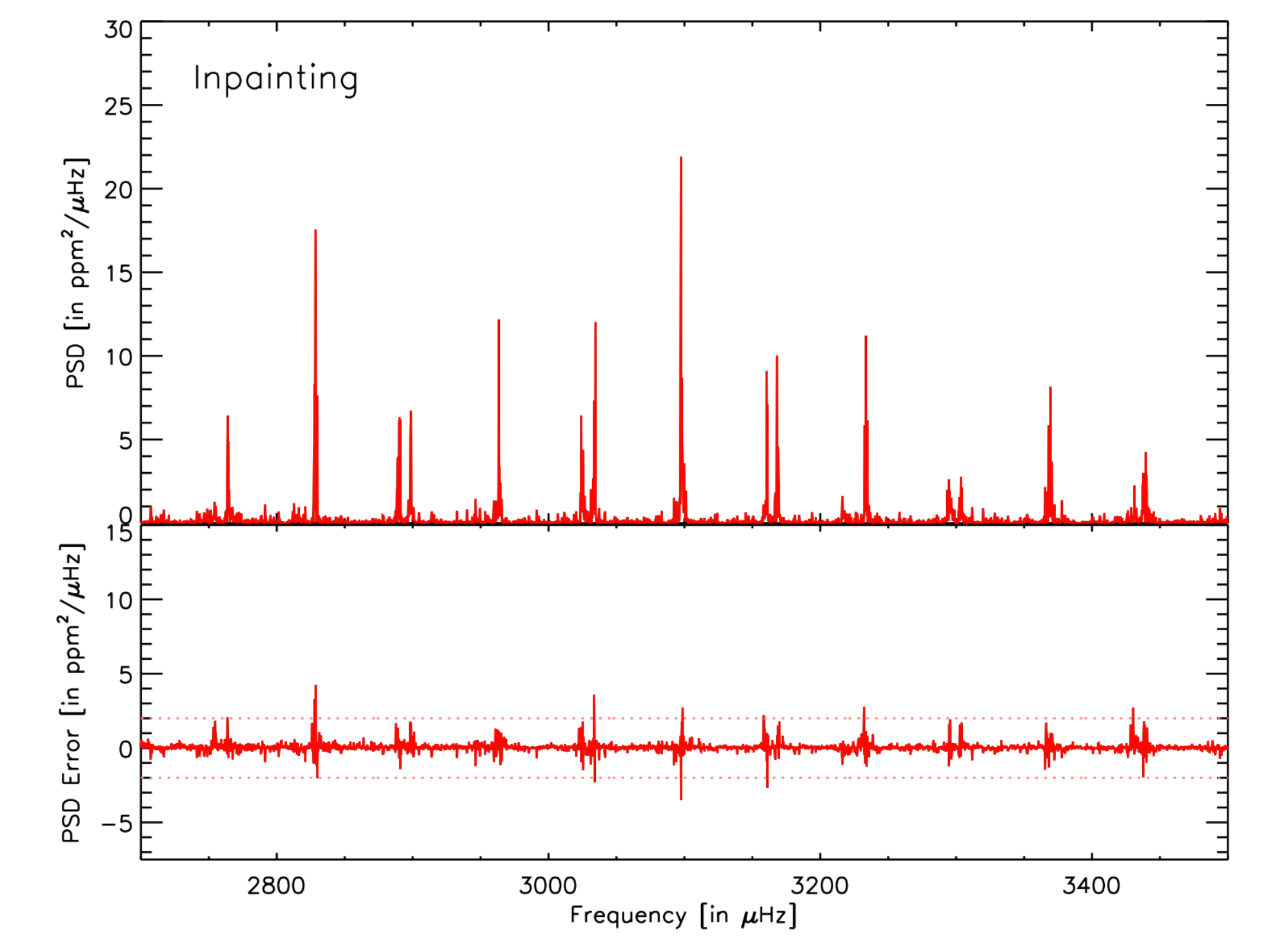}
}
\caption{Power Density Spectrum and errors (in units of ppm$^2/\mu$Hz) estimated from a time series of 50 days with a duty cycle of 50\%. The PDS is computed with a least-squares sinewave fit on the incomplete time series (top panel), using an FFT on the CLEANed time series (middle panel) and using an FFT on the inpainted time series (bottom panel).}
\label{dc50}
\end{figure}

\begin{figure}[htp]
\centerline{
\includegraphics[width=9cm, height=5.9cm]{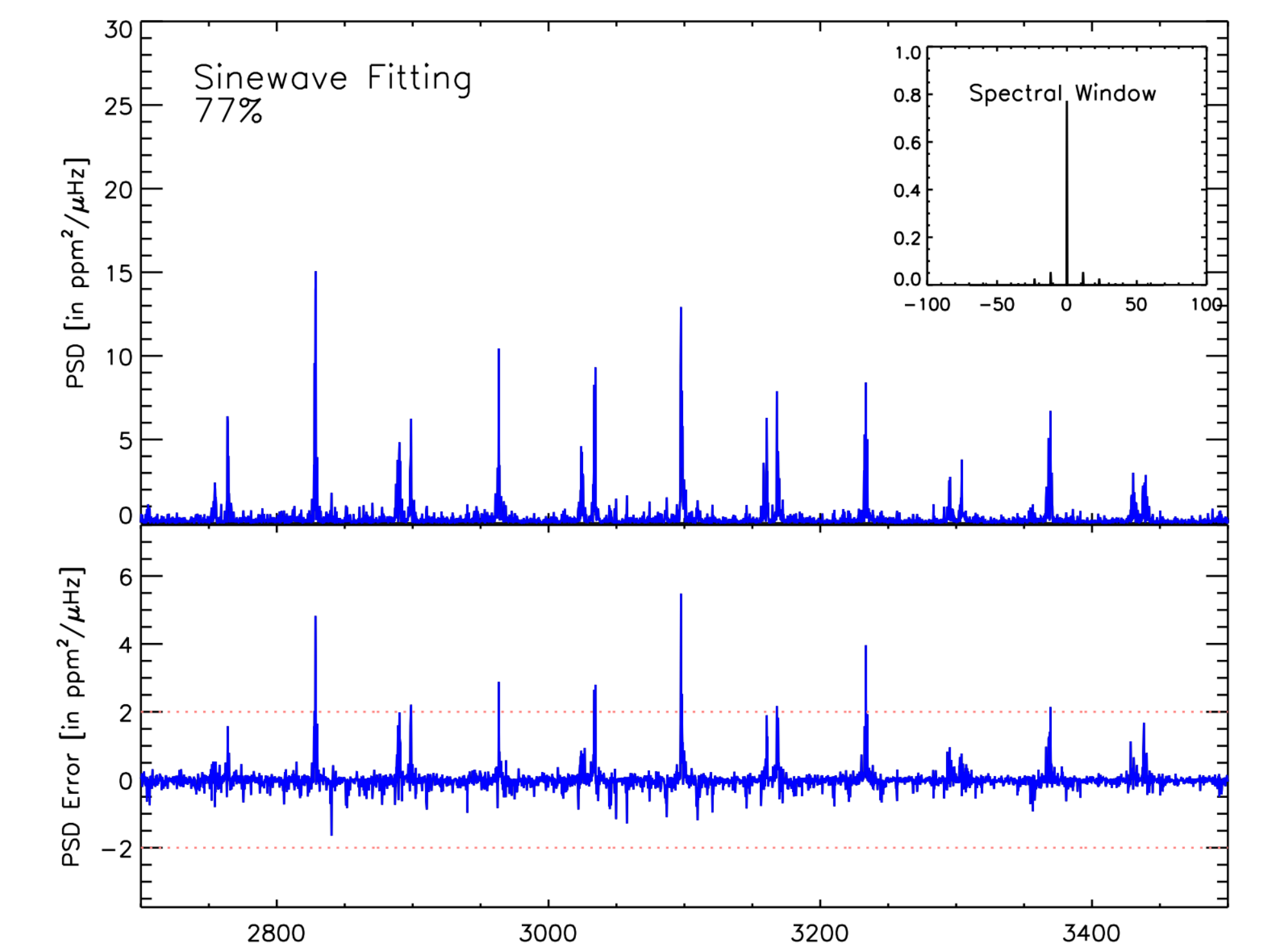}
}
\vspace{0.cm}
\centerline{
\includegraphics[width=9cm, height=5.9cm]{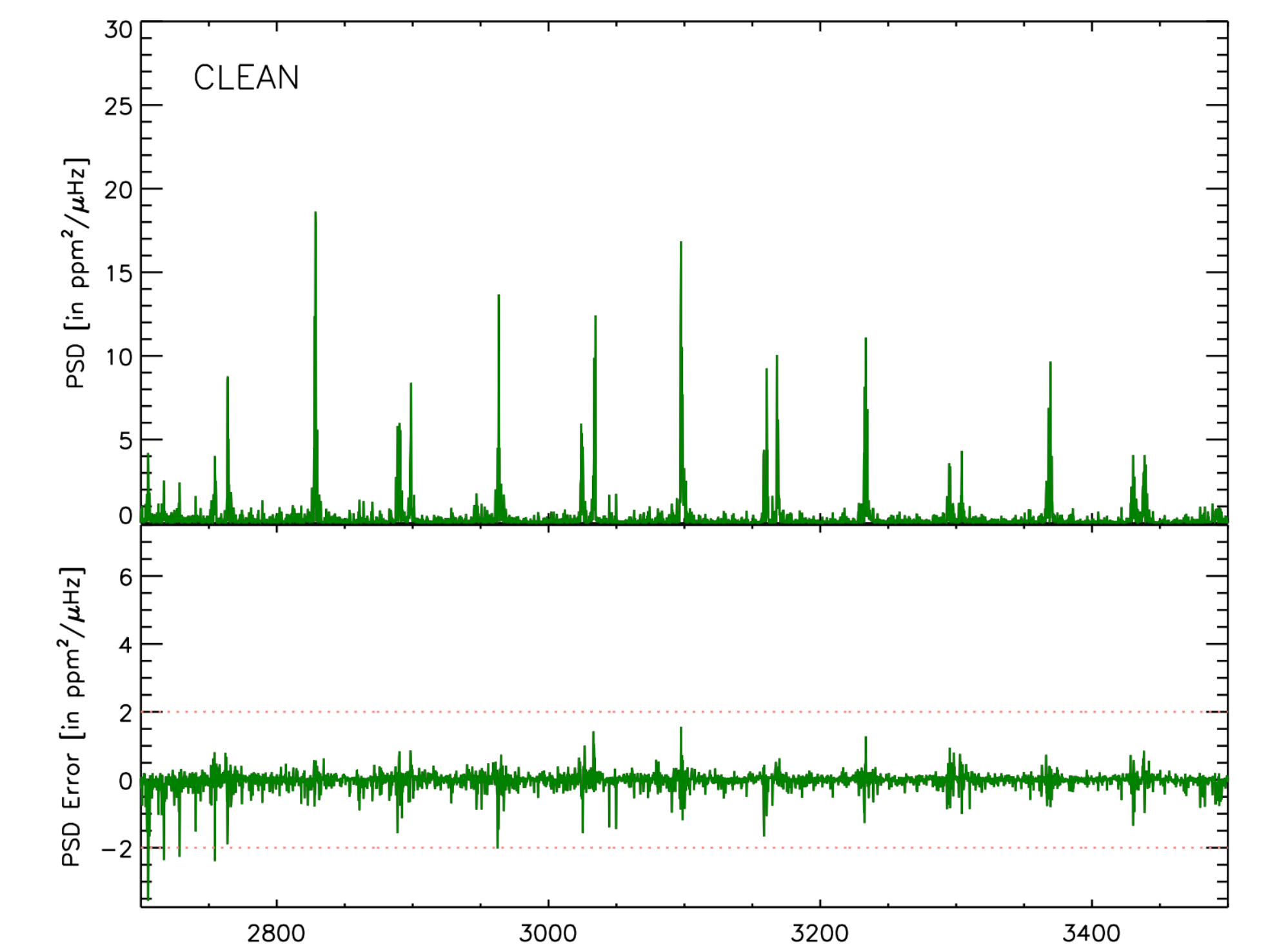}
}
\centerline{
\includegraphics[width=9cm, height=6.3cm]{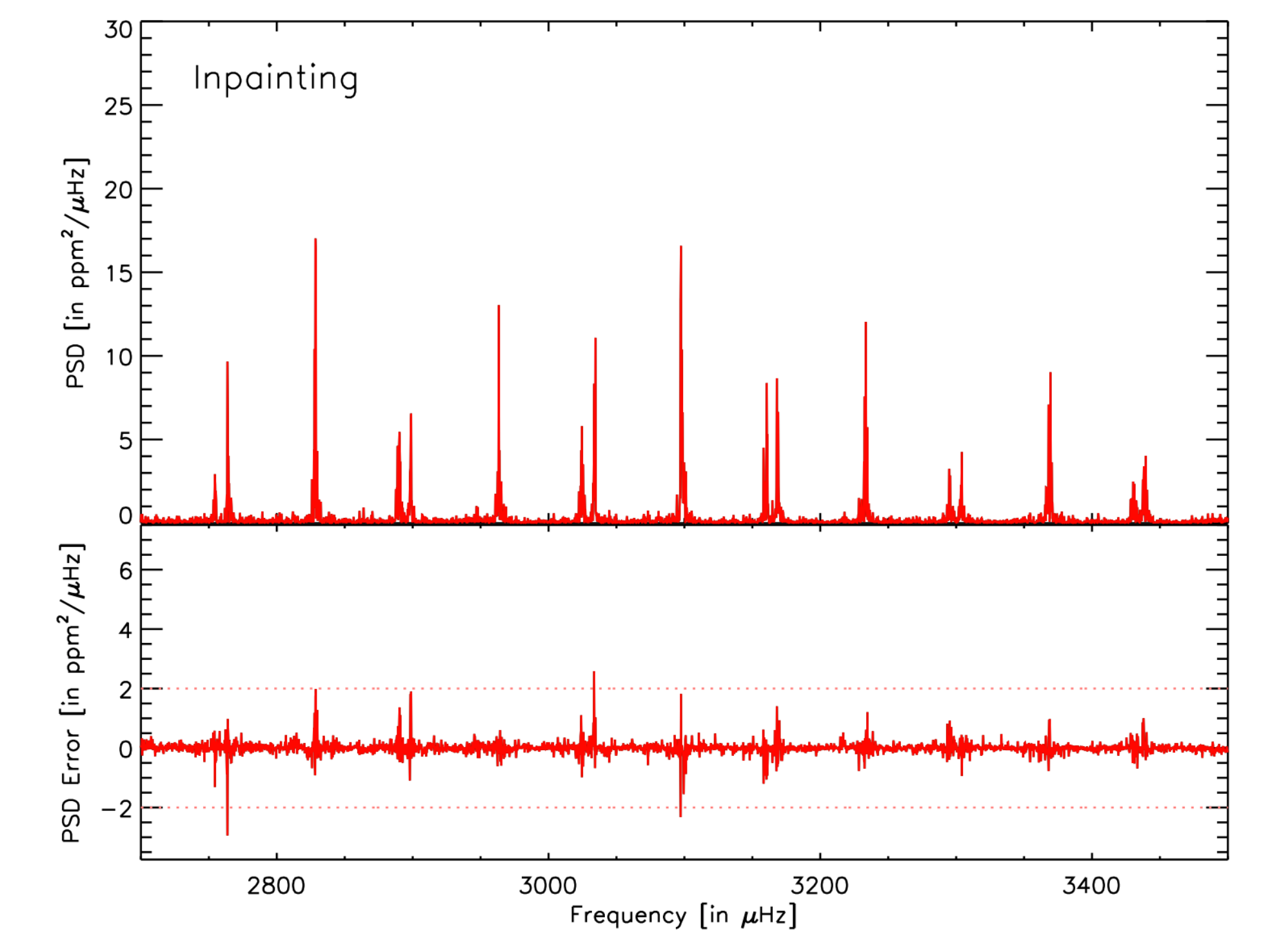}
}
\caption{Power Density Spectrum and errors (in units of ppm$^2/\mu$Hz) estimated from a time series of 50 days with a duty cycle of 77\%. The PDS is computed with a least-squares sinewave fit on the incomplete time series (top panel), using an FFT on the CLEANed time series (middle panel) and using an FFT on the inpainted time series (bottom panel).}
\label{dc77}
\end{figure}

The PDSs obtained with the three methods described in section 2, are shown in Figures~\ref{dc23}, \ref{dc50}, and \ref{dc77} for duty cycles of 23\%, 50\%, and 77\% respectively. In the top panel, the PDS is calculated by the method described in \cite{kepler:frandsen95} which consists of a least-squares sinewave fit (SWF) to the incomplete data and where data points in the gaps are removed from the time series. In the middle panel, the PDS is obtained using a Fast Fourier Transform on a time series reconstructed using the CLEAN algorithm. The bottom panel presents the PDS computed with a Fast Fourier Transform after performing an inpainting. Each panel includes two plots :  the PDS (top) and the error on the PDS (bottom). The plotted error corresponds to the difference between the power spectrum estimated from the complete data and the power spectrum estimated from the incomplete set. Thus positive errors correspond to underestimation of the signal. The magenta dotted lines have been plotted as a reference, they correspond to an error of $\pm$ 0.01 ppm$^2/\mu$Hz. 

For a duty cycle of 23\%, when using the SWF (top panel of Figure~\ref{dc23}) the oscillation modes are convolved by the Fourier transform of the mask, shown in the inset. The amplitude of the oscillation modes is substantially under-estimated due to spectral leakage\footnote[1]{The spectral leakage caused by the window function is  sometimes called incorrectly aliasing. Aliasing should only be used if the spectral leakage is caused by sampling.} of the power into the high amplitude side lobes of the spectral window. Hence, there are several spurious peaks, which correspond to negative values in the PDS error. This leakage is reduced when we increase the duty cycle (thus $T_1$) as expected by Eq. 2. but the spurious peaks are still present in the PDS (top panels of Figures~\ref{dc50} and \ref{dc77}).

With the CLEAN algorithm (middle panel of Figure~\ref{dc23}), we can see the improvement compared to the least-squared sinewave fit. The amplitude of the main high peaks is closer to the original PDS, and the side lobes have been reduced compared to the previous method. However, there are still some side lobes in the power spectrum which correspond to incorrect removal of the peaks. Because of the regularity of the mask, we are in the ``worst case scenario" for the CLEAN algorithm, which looks in an iterative way for the highest peaks in the PDS, assuming they are due to real signal. As seen in Fig. \ref{dc23}, the amplitude of the side lobes is very close to the amplitude of the main lobe. Hence, two side lobes from two nearby main lobes can be larger than the main lobes. This explains the side lobes that remain in the power spectrum.
With duty cycles of 50 and 77\% (middle panels of Figures~\ref{dc50} and \ref{dc77}), the side lobes are significantly reduced similarly to the SWF method but there is still a small spectral leakage even at 77\% duty cycle.

Like CLEAN, applying the inpainting method to the data with 23\% duty cycle (bottom panel of Figure~\ref{dc23}) shows a clear reduction in the side lobes compared to SWF but the amplitude of the peaks is not well recovered. However, there are a few instances where the main peaks are almost completely missed (see large positive peaks in the PDS error plots). These errors are due to the mask not satisfying one of the principal conditions previously laid down ``mask has to be sufficiently random". Also for inpainting this is the ``worst case scenario", because the amount of missing data is significant and the mask is coherent. Even worse, because the mask is regular, it is sparse in the representation $\Phi^T$ of the data (i.e. Multiscale Discrete Cosine Transform) in which the signal is also sparse. In this case, the effect of the window function cannot be fully removed.
The increase of the duty cycle to 50\% and 77\% (bottom panels of Figures~\ref{dc50} and \ref{dc77}) improves the results of the inpainting method where we clearly see the reduction of the side lobes. We also notice that the amplitudes of the main peaks are much better recovered.

In summary, the CLEAN method is better in terms of estimating the amplitude of the modes for typical ground-based data with duty cycles of 23\%, while the inpainting method removes the side lobes better than CLEAN does for the 50\% and 77\% cases.

\subsection{Fitting the acoustic modes of the different Power Density Spectra}
In the case of a 23\% duty cycle, the inpainting method is clearly not optimal to process the regularly gapped ground-based data, but for a duty cycle of 77\%, the inpainting method outperforms the others methods. To further quantify the results obtained in the intermediate case of 50\% duty cycle, we fit the oscillation modes with a maximum likelihood estimation as described in \cite{fit:anderson90, fit:appourchaux98}. Modes are modeled as the sum of Lorentzian profiles \citep[see e.g.][]{fit:kumar88} and are fitted simultaneously over one large separation, i.e. a sequence of $l=2,0,3,$ and 1 modes, assuming a common width for the modes, a common rotational splitting (fixed to $0.4\ \mathrm{\mu Hz}$) and assuming that the Sun is observed from near the equatorial plane of the Sun ($90^{\circ}$ inclination).  We fit five sequences of modes, by fixing within each sequence the visibilities for $l=1,2,3$ modes relatively to the $l=0$ mode \citep[see e.g.][for details on fitting techniques]{fit:ballot11}.

In the PDS obtained by sinewave fitting, the amplitude of the modes is significantly underestimated due to leakage of the power into the side lobes of the spectral window. To minimize the effect of this spectral leakage, the fitted model is convolved with the spectral window.

The PDS reconstructed from the CLEAN method is hard to fit with this technique, since the statistics of the noise is modified in a non trivial way. To 
avoid this problem, we would need to CLEAN the spectrum down to very low amplitude, which is too time consuming.

Fitting results are reported in Table~\ref{tab:fit}. We should remember that the errors are underestimated in the case of the PDS obtained by sinewave fitting, because the statistics used does not take into account the correlation between the points introduced by the window function. Except for $l=3$ modes, we see that the mode parameters are well recovered in both cases. Nevertheless, we must be very cautious by interpreting such a table because it concerned only 5 modes for one given noise realization. Moreover, these results are obtained with good guesses, close to the real parameters. Fits of the PDS obtained by sinewave fitting are very sensitive to the guesses due to the presence of day aliases. This problem is avoided when the series are inpainted  for which side lobes are strongly reduced.

\begin{table}[!htp]
\caption{Fitted parameters: frequency ($\nu$), rms amplitude ($\sigma_{rms}$) and width ($\Gamma$) of modes $l=0$ to $3$ of five radial orders around $\nu_{\rm{max}}$. Fits are done on the PDS obtained by sinewave fitting and the inpainting method for a duty cycle of 50\:\%. As a reference, we use the PDS obtained with the complete series.\label{tab:fit}}
 \begin{tabular}{ccc}
\hline\hline
 DFT 100\:\% &  SWF 50\:\%  & Inpainting 50\:\% \\
\hline
\multicolumn{3}{c}{Frequency $l=0$ [$\mu$Hz]}\\
$2764.09 \pm 0.16$ & $2763.94 \pm 0.14$ & $2764.27 \pm 0.19$ \\
$2898.53 \pm 0.18$ & $2898.74 \pm 0.11$ & $2898.61 \pm 0.17$ \\
$3033.76 \pm 0.17$ & $3033.83 \pm 0.14$ & $3033.74 \pm 0.17$ \\
$3168.65 \pm 0.20$ & $3168.61 \pm 0.16$ & $3168.37 \pm 0.15$ \\
$3304.08 \pm 0.31$ & $3304.01 \pm 0.27$ & $3303.59 \pm 0.33$ \\
\hline 
\multicolumn{3}{c}{Frequency $l=1$ [$\mu$Hz]}\\
$2828.29 \pm 0.14$ & $2828.23 \pm 0.09$ & $2828.47 \pm 0.21$ \\
$2963.33 \pm 0.58$ & $2963.84 \pm 0.12$ & $2963.15 \pm 0.20$ \\
$3098.13 \pm 0.16$ & $3097.93 \pm 0.10$ & $3097.95 \pm 0.14$ \\
$3233.39 \pm 0.24$ & $3233.50 \pm 0.28$ & $3233.86 \pm 0.16$ \\
$3368.78 \pm 0.21$ & $3368.85 \pm 0.15$ & $3368.92 \pm 0.19$ \\
\hline 
\multicolumn{3}{c}{Frequency $l=2$ [$\mu$Hz]}\\
$2754.26 \pm 0.21$ & $2754.19 \pm 0.16$ & $2753.44 \pm 0.38$ \\
$2889.70 \pm 0.15$ & $2889.95 \pm 0.09$ & $2889.84 \pm 0.12$ \\
$3024.71 \pm 0.19$ & $3024.73 \pm 0.19$ & $3025.04 \pm 0.22$ \\
$3159.81 \pm 0.13$ & $3159.97 \pm 0.14$ & $3160.01 \pm 0.11$ \\
$3295.44 \pm 0.34$ & $3295.54 \pm 0.31$ & $3295.37 \pm 0.28$ \\
\hline 
\multicolumn{3}{c}{Frequency $l=3$ [$\mu$Hz]}\\
$2813.19 \pm 0.35$ & $2813.21 \pm 0.25$ & $2813.17 \pm 0.22$ \\
$2946.69 \pm 0.18$ & $2946.93 \pm 0.25$ & $2947.53 \pm 0.11$ \\
$3082.11 \pm 0.20$ & $3081.95 \pm 0.34$ & $3080.49 \pm 0.44$ \\
$3216.65 \pm 0.34$ & $3217.29 \pm 0.35$ & $3217.54 \pm 0.07$ \\
$3352.38 \pm 0.64$ & $3353.29 \pm 1.08$ & $3352.30 \pm 0.80$ \\
\hline
\multicolumn{3}{c}{Amplitude $\sigma_{\mathrm{rms}}$ [ppm] $l=0$}\\
$  3.21 \pm  0.30$ & $  3.16 \pm  0.20$ & $  2.84 \pm  0.24$ \\
$  3.84 \pm  0.35$ & $  3.43 \pm  0.23$ & $  3.80 \pm  0.34$ \\
$  3.89 \pm  0.30$ & $  3.70 \pm  0.20$ & $  3.65 \pm  0.27$ \\
$  3.76 \pm  0.31$ & $  3.34 \pm  0.19$ & $  4.51 \pm  0.47$ \\
$  2.88 \pm  0.19$ & $  2.69 \pm  0.14$ & $  2.74 \pm  0.19$ \\
\hline
\multicolumn{3}{c}{Width $\Gamma$ [$\mu$Hz]  $l=0$}\\
$  0.66 \pm 0.18$ & $  0.65 \pm 0.21$ & $  0.95 \pm 0.24$ \\
$  0.70 \pm 0.17$ & $  0.56 \pm 0.15$ & $  0.53 \pm 0.12$ \\
$  0.84 \pm 0.17$ & $  0.91 \pm 0.18$ & $  0.91 \pm 0.17$ \\
$  0.90 \pm 0.19$ & $  0.94 \pm 0.19$ & $  0.36 \pm 0.09$ \\
$  1.61 \pm 0.31$ & $  1.37 \pm 0.27$ & $  1.45 \pm 0.29$ \\
\hline
 \end{tabular}
\end{table}

\section{Application to realistic simulations of Space-Based data}
\label{space}
Space-based observations in particular by CoRoT and {\it Kepler}, have duty cycles around 90\%.
However, even small gaps on the time series can introduce significant artifacts in the power spectrum if they are regularly distributed.
\subsection{Corot-like data}

The CoRoT mission provides 3 to 5 month-long observations of high-precision photometry.
However, these time series are periodically perturbed by high-energy particles hitting the satellite when crossing the South Atlantic Anomaly (SAA) \citep[e.g.][]{corot:auvergne09} resulting in gaps in the data. The gaps in the CoRoT light curves have a typical time duration of 20 min and a periodicity that originates from the orbital period of the satellite. Fortunately, even if the satellite is crossing the SAA regularly, the perturbation is not the same for each orbit. Thus, the observational window of CoRoT is more complex and can not be easily modeled. We can still estimates the spectral window using a Fast Fourier Transform. 
The inset in the top panel of Fig. \ref{dc_corot} shows a typical spectral window of CoRoT. There is about 15\% of the power that leaks from the main lobe into the side lobes. However, in contrast with the previous masks, the side lobes are not significant. In this case, only 0.1\% of the amplitude leaks into the first side lobes because the power is spread into more frequencies. It would be 3\%, if  the mask was regular.

\subsubsection{Oscillation modes}

In this section, we study the impact of these small repetitive gaps on the detection and estimation of the oscillation modes.
For this purpose, we consider, the observation window of the star HD~169392 \citep{corot:mathur13} which was observed by CoRoT over 91.2 continuous days during the third long-run in the galactic centre direction with a duty cycle of 83.4\% and with a sampling of 32s.
This window is applied to 91.2 days of VIRGO data in which the original CoRoT 32s cadence has been converted into the 60s cadence of VIRGO data.
The bottom panel of Fig. \ref{original_psd} shows the Power Density Spectrum (PDS) estimated from 91.2 days of observations with a complete time coverage (duty cycle of 100\%).

\begin{figure}[htp]
\centerline{
\includegraphics[width=9cm, height=5.9cm]{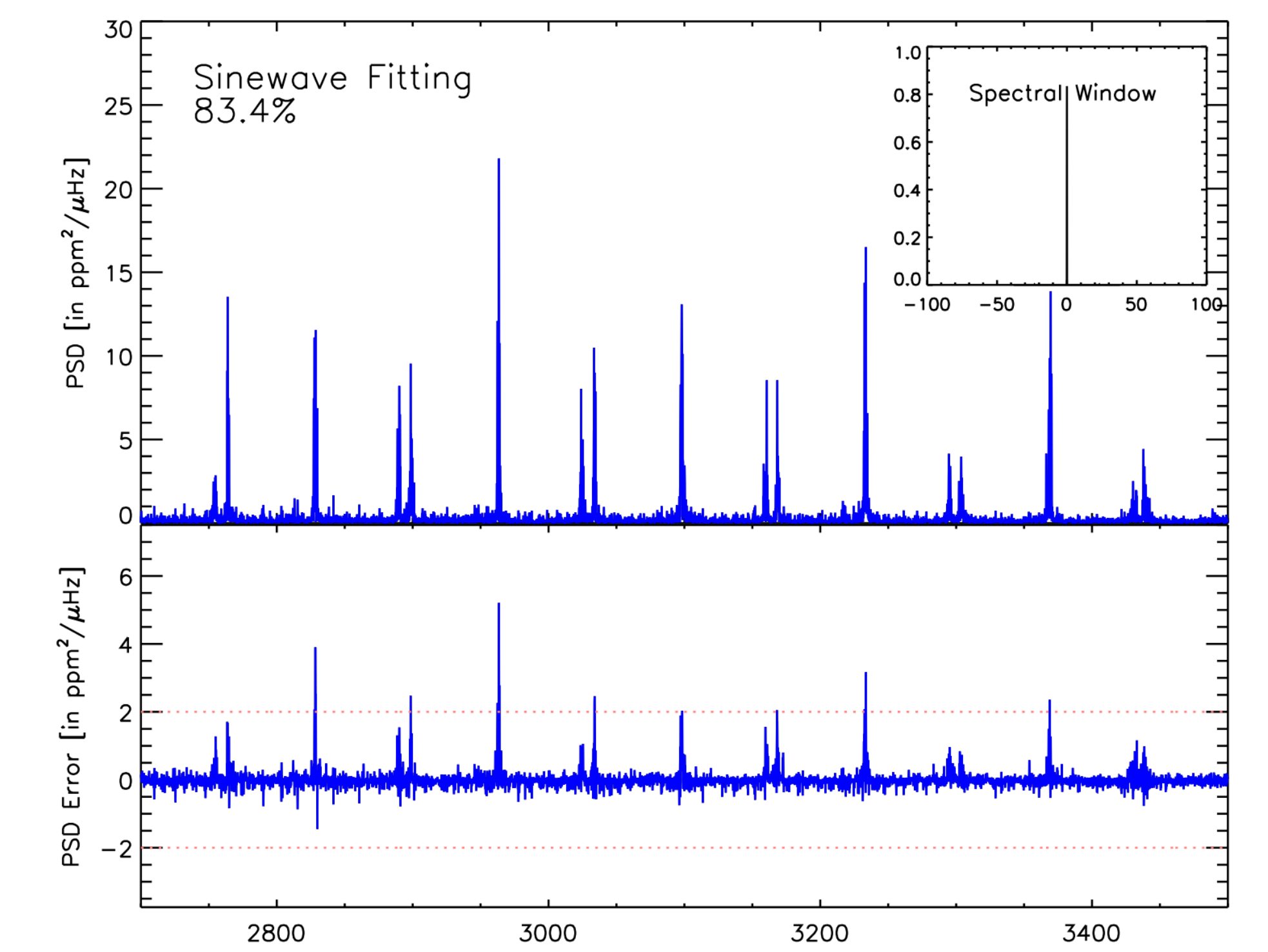}
}
\vspace{0.cm}
\centerline{
\includegraphics[width=9cm, height=5.9cm]{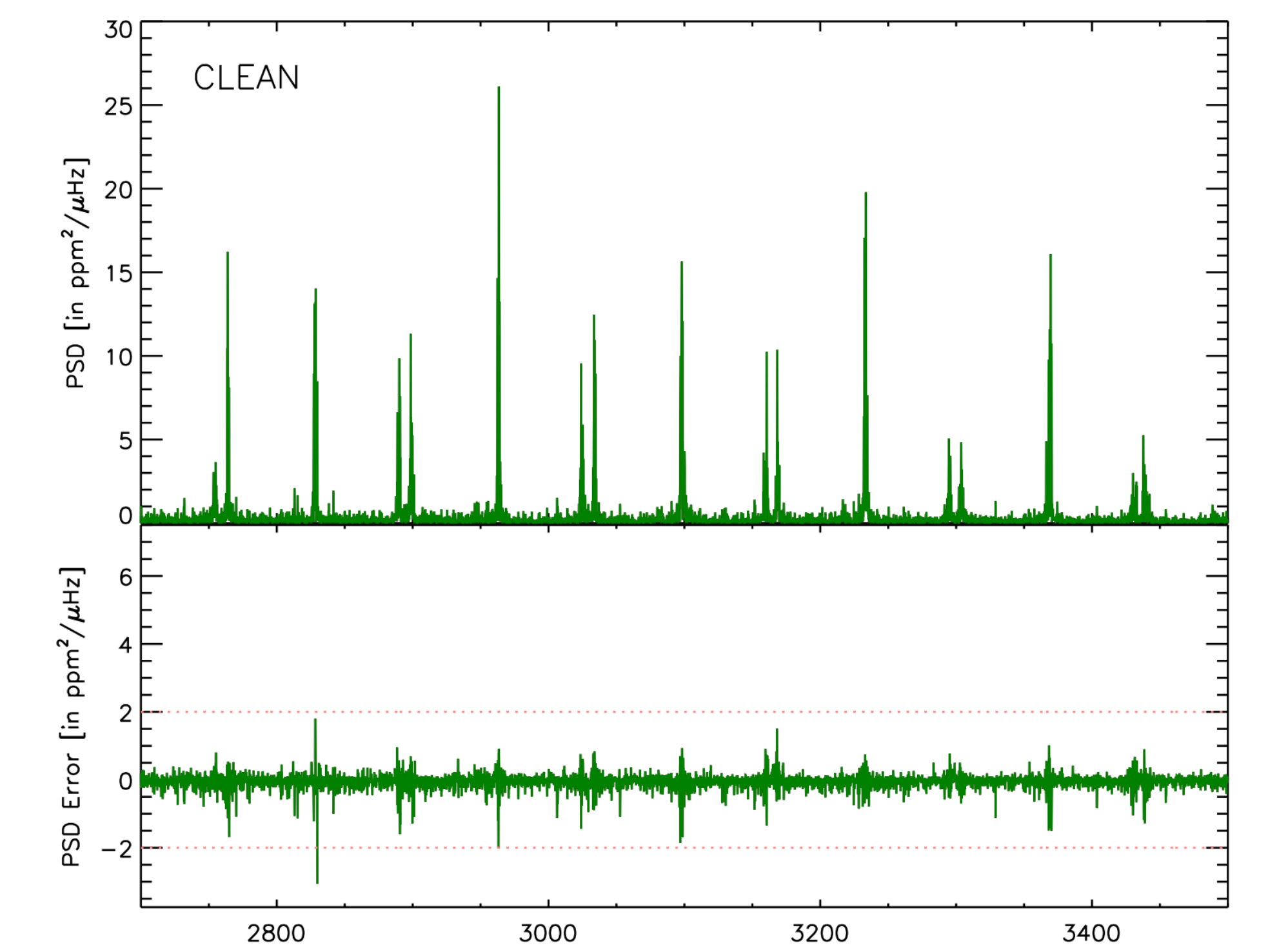}
}
\centerline{
\includegraphics[width=9cm, height=6.3cm]{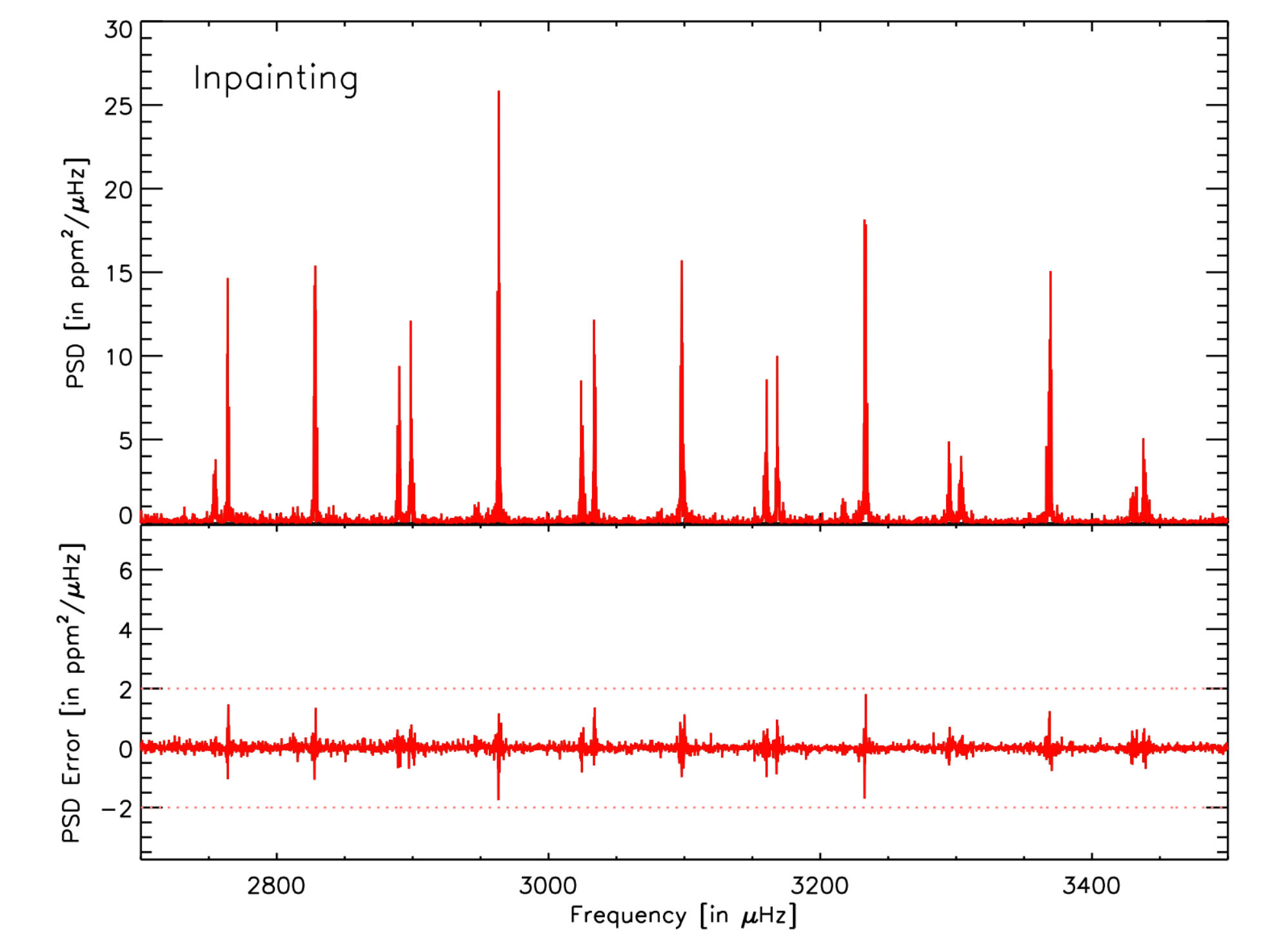}
}
\caption{Power Density Spectrum and errors (in units of ppm$^2/\mu$Hz) estimated from a time series of 91.2 days masked with a mask of Corot corresponding to a duty cycle of 83.4\%. The PDS is computed with a least-squares sinewave fit on the incomplete time series (top panel), using an FFT on the CLEANed time series (middle panel) and using an FFT on the inpainted time series (bottom panel).}
\label{dc_corot}
\end{figure}

Figure \ref{dc_corot} shows the results for this CoRoT-like data following the previous format.
As expected from the shape of the window function in the Fourier domain (see inset), the side lobes have almost disappeared from the PDS obtained by sinewave fitting.
However, there is still a small spectral leakage of the power into the side lobes as said previously.
The PDS estimated from CLEAN yields a good estimation of the amplitude of the oscillation modes and the side lobes have almost disappeared. This is due to the observation window that produces a low level of side lobes, that helps the CLEAN algorithm to avoid detecting false peaks.
Once more, there is a significant improvement in the estimation of the amplitude of the oscillation modes for the inpainted data.
This is certainly due to the random way in which the high-energy particles hit the satellite. This insures a given amount of incoherence to the mask despite the regularity with which the satellite crosses the SAA area.

\subsubsection{Background power spectrum}

Again, we study the impact of the small repetitive gaps induced by the SAA crossing, on the power spectrum estimation.
However, this time we focus on the background part of the power spectrum rather than the oscillation modes.

In addition to the oscillation modes, the asteroseismic observations allows us to measure the stellar granulation signature by the characterization of the background power spectrum \citep{granulation:mathur11}.
The spectral signature of granulation is expected to reveal time scales characteristic of the convection process in the stars.
However, this analysis requires that we estimate the full power spectrum.

For this study, the CLEAN algorithm has not been considered because it is very time-consuming to reconstruct the full power spectrum as it is explained in section \ref{time processing}.

\begin{figure}[htp]
\centerline{
\includegraphics[width=9cm]{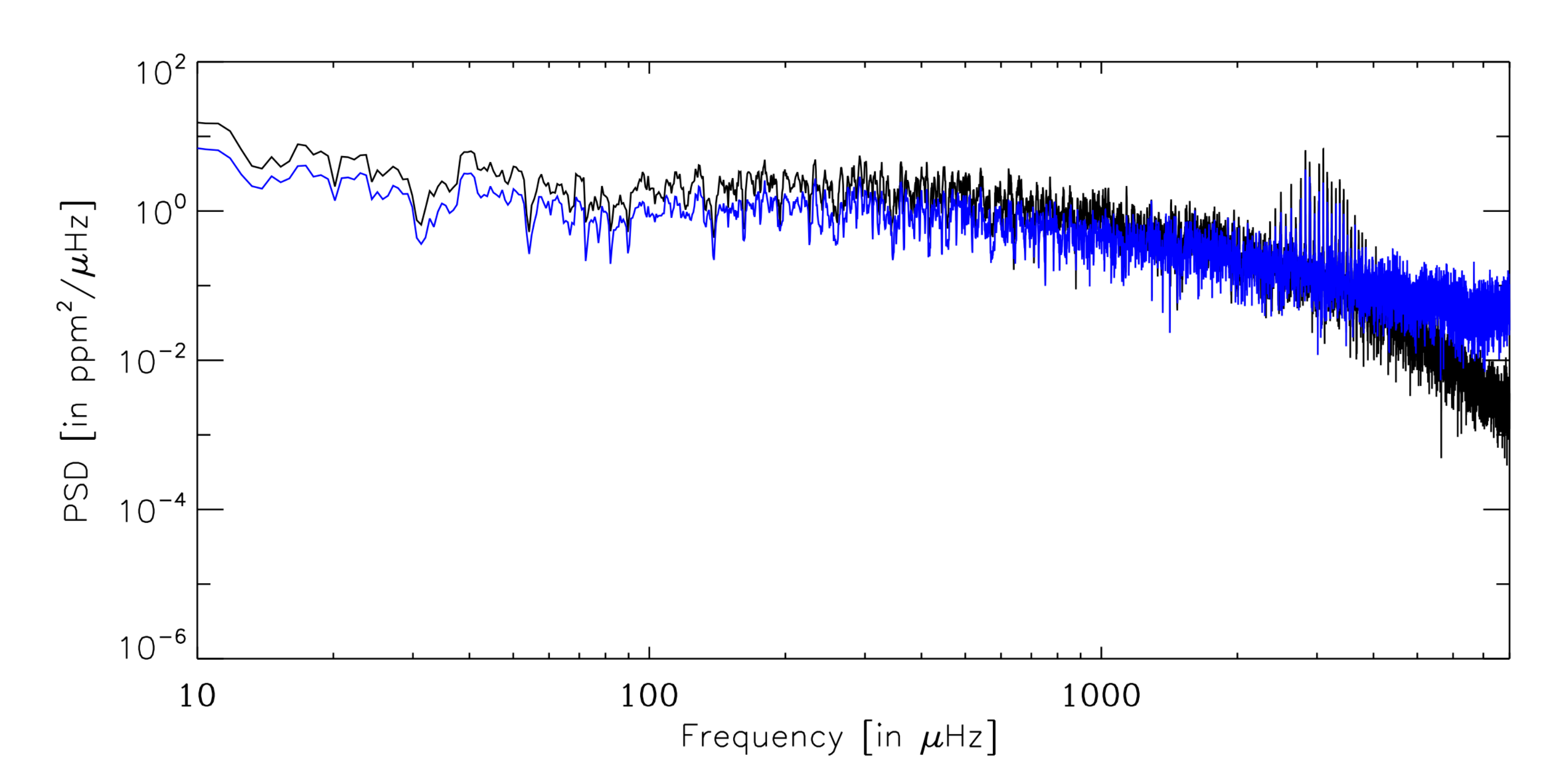}
}
\centerline{
\includegraphics[width=9cm]{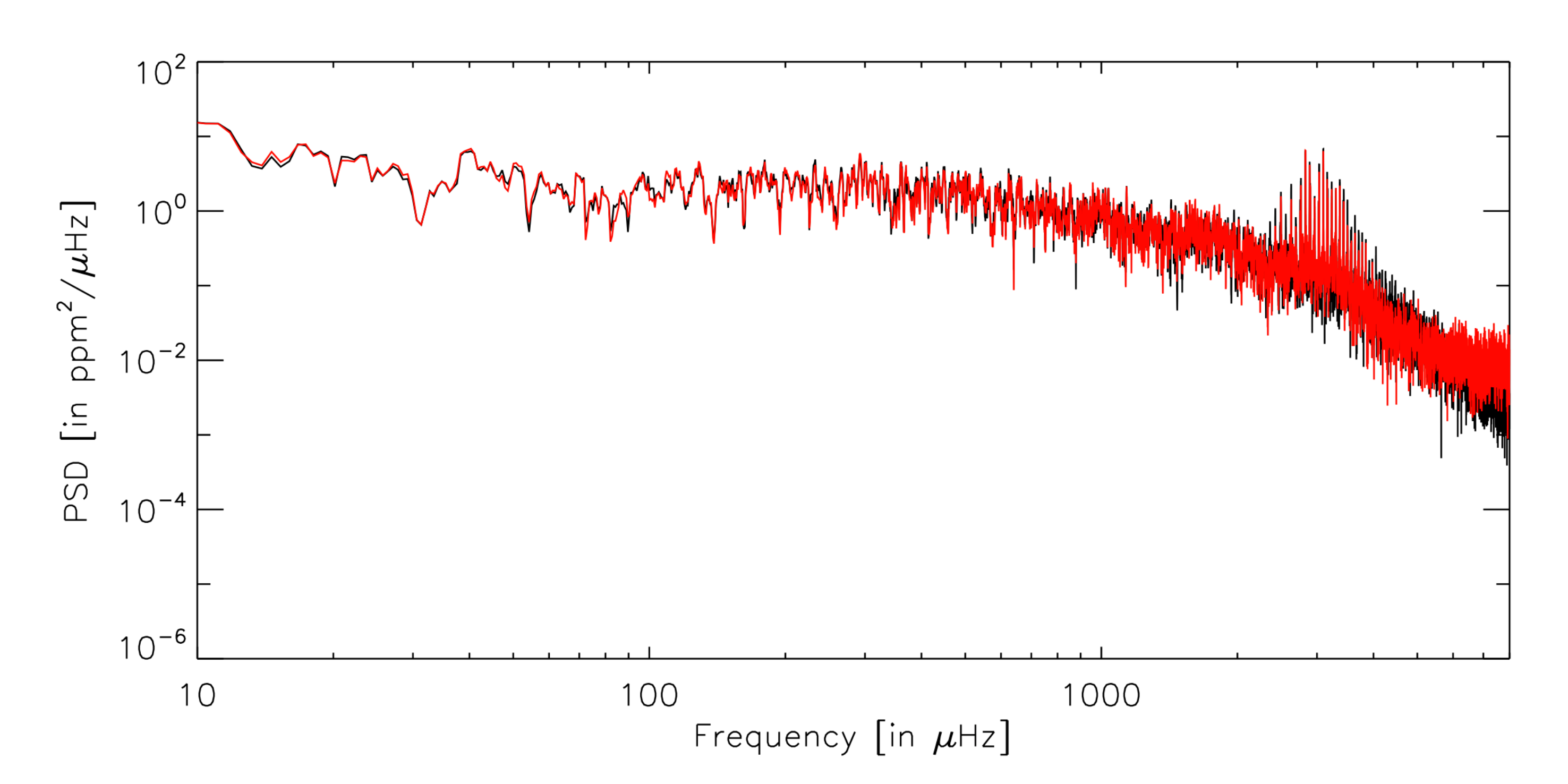}
}
\caption{Power Density Spectrum smoothed with a 4 points boxcar function (in units of ppm$^2/\mu$Hz) estimated from a time series of 24 days of VIRGO/SPM masked with a mask of CoRoT with a duty cycle of 70\%. The black curve corresponds to the original PDS estimated from the complete data. In the top panel, the blue curve corresponds to the PDS calculated with a least-squares sinewave fit on the incomplete data. Bottom panel, the red curve correspond to the PDS estimated using a Fast Fourier Transform on the inpainted data.}
\label{background}
\end{figure}

Top panel of Fig. \ref{background} shows the original PDS of VIRGO/SPM estimated from the complete data compared to a PDS estimated from the incomplete data.
We can see that the level of power at high frequencies is over-estimated, resulting in a bad granulation characterization. Moreover, the increase in the high-frequency noise affects the oscillation modes at high-frequency \citep{golf:jimenez11} as well as the pseudo-mode region \citep{golf:garcia98}.
In the bottom panel of Fig. \ref{background}, we show the power spectrum estimated after inpainting of the data and it appears that the inpainting technique recovers better the high-frequency level of the power spectrum.


To quantify the difference between the backgrounds of the different PDS, we fit them with the following model:
\begin{eqnarray}\label{eq:bg}
B(\nu) &=& \frac{\zeta_{\mathrm{g}}\sigma^2_{\mathrm{g}}}{1+(2\pi\tau_{\mathrm{g}}\nu)^{3.5}} +
\frac{\zeta_{\mathrm{f}}\sigma^2_{\mathrm{f}}}{1+(2\pi\tau_{\mathrm{f}}\nu)^{6.2}} + \nonumber \\
&&\exp \left[ -\frac{(\nu-\nu_o)^2}{2\delta_o^2}\right] + W.
\end{eqnarray}
The first two terms are Harvey profiles. 
The first component is to model the granulation and is parametrized with $\tau_{\mathrm{g}}$, the characteristic time scale of the granulation, $\sigma_{\mathrm{g}}$ its rms amplitude. $\zeta_{\mathrm{g}}=2.73$ is a normalisation factor. 
The second component, possibly originated from faculae is parametrized with  $\tau_{\mathrm{f}}$ and $\sigma_{\mathrm{f}}$. The normalisation factor is $\zeta_{\mathrm{f}}=3.00$. The exponent 3.5 and 6.2 are chosen according to \cite{feculae:karoff12}. 
The third component is a Gaussian function parametrized with $\nu_o$ and $\delta_o$ to account for the power excess due to oscillation modes. Finally, $W$ is the white noise component.
Spectra are fitted using a maximum likelihood estimation between $\mathrm{100\ \mu Hz}$ and the Nyquist frequency, by assuming that the 
noise follows a $\chi_2^2$ distribution.

\begin{table}
\begin{center}
\begin{tabular}{lccccc}
\hline
\hline
           & $\tau_{\mathrm{g}}$ & $\sigma_{\mathrm{g}}$ & $\tau_{\mathrm{f}}$ & $\sigma_{\mathrm{f}}$ & $W$\\
           &        [s]       &     [ppm]           &        [s]       &     [ppm]          &[$\frac{10^{-3}\text{ppm}^2}{\mu \text{Hz}}$]\\
\hline
\hline
Original   &   $190\pm 7$     & $42\pm 1$           &   $48\pm 1$     & $30\pm 1$           & $1.34 \pm 0.12$ \\
SWF        &  $273\pm 24$     & $23\pm 1$           &   $94\pm 4$     & $23\pm 1$           & $59.5 \pm 1.1$ \\
Inpaint. &   $200\pm 9$     & $40\pm 1$           &   $57\pm 1$     & $30\pm 1$           & $8.78 \pm 0.22$ \\
\hline
\end{tabular}
\end{center}
\caption{Fitted parameters for the background (see eq.~\ref{eq:bg}).}
\label{bgtable}
\end{table}

Table~\ref{bgtable} shows the results of the fitting. In the spectrum obtained with the sinewave fitting, the Harvey parameters are badly fitted. 
By contrast, the inpainted spectrum allows us to correctly recover both the amplitude and the characteristic time scale for the two components. 
The main visible change is an increase of the white noise component, generated by an increase of the high frequency noise.
However, this excess of high frequency noise remains small compared to the SWF case.

\subsubsection{Processing time}
\label{time processing}
For space-based data for which we have to deal with long and continuous observations, the impact of the gaps is smaller, because the duty cycle is approaching 100\%.
For this reason, the way the missing data are handled is frequently driven by the time it takes to do the correction.  

We have estimated the processing time for the 3 methods used in this paper:
\begin{itemize}
\item The sinewave fit (SWF) has a time complexity of $O(N^2)$ where N is the number of data points of the time series.
This time complexity can be reduced to $O(N N_f)$ if we are just interested in reconstructing a number of frequencies $N_f$.

\item The CLEAN algorithm has a time complexity of $O(N^3)$ 
In the same way, it can be reduced to $O(N N_f^2)$ if we are just interested in a part of the power spectrum.
This scaling means that for long time series, this algorithm may prove to be very time-consuming. 

\item The inpainting algorithm has a time complexity of $O(I_{max} N\log(N))$.  
The number of iterations $I_{max}$ is taken equal to $100$.
\end{itemize}
For a time series of 50 days observed with a sampling of 32s, which is the case for the CoRoT long-cadence observations, with a duty cycle of about $77\%$, it takes 4 hours to compute one tenth of the full SWF power spectrum up to the Nyquist frequency, about 4 days to compute one tenth of the CLEAN power spectrum and only 4 min to compute the full inpainted power spectrum on a 2 x 2.4GHz Intel Xeon Quad-Core processor.
Thus, even considering just a fraction of the power spectrum, the inpainting algorithm is 1000 times faster than the CLEAN algorithm and 60 times faster than the SWF method.

\subsection{{\it Kepler} -like data}
Since its launch in May 2009, NASA's {\it Kepler} mission  \citep{kepler:borucki10} has also been used to study stars with asteroseismology.
Indeed, around 2000 solar-like stars \citep{kepler:chaplin11} have been observed continuously in short cadence --with a sampling rate of $\sim$\,58.85s \citep[see for details, ][]{kepler:gilliland10b,kepler:garcia11} - for at least 1 month, and up to 2 years.
  In addition, around 15,000 red giants showing solar-like oscillations have been observed in long cadence (sampling of $\sim$\,29.5 min) \citep[e.g.][]{kepler:bedding10,kepler:huber11,kepler:stello13}.


{\it Kepler} data typically reaches duty cycles of about 93 \%, where gaps are mainly due to a combination of data downlinks (of around one day in average) every three months and desaturations  (of one long-cadenced data point and several short-cadenced data points) every 3 days \citep{kepler:christiansen13}.

Due to their regularity these single-point gaps produce a similar effect at high frequencies as those showed for CoRoT in Figure 7, which is most pronounced for stars showing high-amplitude modulation at low frequencies. Hence, by only inpainting the single long-cadence point missing every three days, we could almost completely remove the spectral leakage of power at high frequencies in the {\it Kepler} data.

\subsubsection{Irregularly sampled data}
The inpainting algorithm has not yet been developed for irregularly sampled data, such as {\it Kepler} data, because the Multiscale Discrete Cosine Transform that is used assumes regularly sampled data. Therefore, before applying the algorithm we place the irregularly Kepler data into a regular grid using the nearest neighbor resampling algorithm \citep{gaps:broersen09}. We therefore built a new time series with a sampling rate equal to the median of the original. Each point of the new series is built from the closest observation (irregularly sampled) if it is within half of the new sampling distance. The regular grid point is set to zero if there is no original observation falling within the new grid \citep[for further details see][]{gaps:garcia14}.

\section{Conclusions}

All improvements in the gap-filling data are of special importance for the analysis of asteroseismology data.
We have shown in this paper, that CLEAN and inpainting are efficient methods to deal with gaps in ground-based and spaced-based data.
Both CLEAN and inpainting methods are based on iterative algorithms that try to deconvolve the observed time series from the window function. However, the difference comes from the way the deconvolution is conducted. In the CLEAN algorithm a direct deconvolution is performed for each frequency. In the inpainting algorithm, the window function is deconvolved indirectly by filling in the gaps.

We showed that CLEAN allows us to process ground-based data with low duty-cycles (lower than 50\%) more efficiently than the inpainting method as it better recovers the amplitudes of the modes. In addition, these low duty-cycle time series can be tackled with a reasonable amount of time by CLEAN that removes the data points from the gaps.
For duty cycles larger than 50\%, the inpainting becomes more interesting. It provides similar and even slightly better results than CLEAN but for a much shorter computational time. Indeed for 50 days time series with a sampling of 32 s and a duty cycle of 77\%, it is about 1000 times faster than CLEAN.

For data with higher duty cycles, typical of space-based observations, the inpainting algorithm retrieves the modes better and becomes a powerful tool that we can apply to thousands of stars in a short amount of time. This is very important in the framework of the {\it Kepler} mission that provided very long time series (around 4 years) for hundred of thousands of stars covering the HR diagram  \citep[e.g.][]{kepler:gilliland10a}.

Finally, we showed that it is important to fill the gaps of the data to better characterize the background at high frequency as well as the granulation components, which are quite affected. Given the computation time factor, the choice of the method would lean towards the inpainting because it would be much more time consuming to compute the whole power spectrum with CLEAN.



The {\it Kepler}  Asteroseismic Scientific Consortium has developed its own automated correction software \citep{kepler:garcia11} to correct most of the light curves with a minimum human intervention. As a result of this study, the inpainting technique has been added to the pipeline to improve the asteroseismic studies.

Furthermore, the inpainting software presented in this study is now publicly available\footnote{The \emph{Kepler}-inpainting software can be found in \url{http://irfu.cea.fr/Sap/en/Phocea/Vie_des_labos/Ast/ast_visu.php?id_ast=3346}}.

\begin{acknowledgements}
The research leading to these results has received funding from the European Community's Seventh Framework Programme (FP7/2007-2013) under grant agreement no. 269194 (IRSES/ASK) and ERC-228261 (SparseAstro). R.A.G, S.M, and S.P. acknowledge the School of Physics staff of the University of Sydney for their warm hospitality during the IRSES research program. S.M. thanks the University of Sydney for their travel support. NCAR is partially funded by the National Science Foundation. This work was partially supported by the NASA grant NNX12AE17G.
\end{acknowledgements}



\bibliographystyle{aa} 
\bibliography{BIBLIO_sav,inpainting1d}

\end{document}